%% file: sample-acmlarge.tex
\documentclass[acmlarge]{acmart}

\usepackage{booktabs} 
\usepackage{multirow}
\usepackage{eucal}
\usepackage[ruled]{algorithm2e} 
\usepackage{adjustbox}
\usepackage{diagbox}
\usepackage{hyperref}
\usepackage{bookmark}
\makeatletter
\newcommand*{\rom}[1]{\expandafter\@slowromancap\romannumeral #1@}
\makeatother

\usepackage{subfig}
\usepackage{xspace}
\graphicspath{{Figures/}}

\newcommand{\paratitle}[1]{\vspace{1ex}\noindent \textbf{#1}}

\SetAlFnt{\small}
\SetAlCapFnt{\small}
\SetAlCapNameFnt{\small}
\SetAlCapHSkip{0pt}
\IncMargin{-\parindent}

\acmJournal{CSUR}
\acmVolume{1}
\acmNumber{1}
\acmArticle{1}
\acmYear{2018}
\acmMonth{7}
\acmArticleSeq{1}

\setcopyright{usgovmixed}

\acmDOI{0000001.0000001}

\begin{document}
\title{Deep Learning based Recommender System: A Survey and New Perspectives}
\author{Shuai Zhang}
\orcid{1234-5678-9012}
\affiliation{%
  \institution{University of New South Wales}
  \streetaddress{K17, CSE, UNSW}
  \city{Sydney}
  \state{NSW}
  \postcode{2052}
    \country{Australia}
}
\email{shuai.zhang@unsw.edu.au}

\author{Lina Yao}
\affiliation{%
  \institution{University of New South Wales}
  \streetaddress{K17, CSE, UNSW}
  \city{Sydney}
  \state{NSW}
  \postcode{2052}
  \country{Australia}
}
\email{lina.yao@unsw.edu.au}

\author{Aixin Sun}
\affiliation{%
  \institution{Nanyang Technological University}
  \country{Singapore}
}
\email{axsun@ntu.edu.sg}

\author{Yi Tay}
\affiliation{%
  \institution{Nanyang Technological University}
  \country{Singapore}
}
\email{ytay017@e.ntu.edu.sg}

\begin{abstract}

With the ever-growing volume of online information, recommender systems have been an effective strategy to overcome such information overload. The utility of recommender systems cannot be overstated, given its widespread adoption in many web applications, along with its potential impact to ameliorate many problems related to over-choice. In recent years, deep learning has garnered considerable interest in many research fields such as computer vision and natural language processing, owing not only to stellar performance but also the attractive property of learning feature representations from scratch. The influence of deep learning is also pervasive, recently demonstrating its effectiveness when applied to information retrieval and recommender systems research. Evidently, the field of deep learning in recommender system is flourishing. This article aims to provide a comprehensive review of recent research efforts on deep learning based recommender systems. More concretely, we provide and devise a taxonomy of deep learning based recommendation models, along with providing a comprehensive summary of the state-of-the-art. Finally, we expand on current trends and provide new perspectives pertaining to this new exciting development of the field.




\end{abstract}

%
%




%
%

\thanks{Yi Tay is added as an author later to help revise the paper for the major revision.}
\thanks{Author's addresses: S. Zhang {and} L. Yao, University of New South Wales; emails: shuai.zhang@unsw.edu.au; lina.yao@unsw.edu.au; A. Sun {and} Y. Tay, Nanyang Technological University; email: axsun@ntu.edu.sg; ytay017@e.ntu.edu.sg;}
\settopmatter{printfolios=true}
\maketitle

\renewcommand{\shortauthors}{S. Zhang et al.}

\input{survey}

\end{document}

%% file: survey.tex
\section{Introduction}
Recommender systems are an intuitive line of defense against consumer over-choice. Given the explosive growth of information available on the web, users are often greeted with more than countless products, movies or restaurants. As such, personalization is an essential strategy for facilitating a better user experience. All in all, these systems have been playing a vital and indispensable role in various information access systems to boost business and facilitate decision-making process~\cite{jannach2010recommender,ricci2015recommender} and are pervasive across numerous web domains such as e-commerce and/or media websites.


In general,  recommendation lists are generated based on user preferences, item features, user-item past interactions and some other additional information such as temporal (e.g., sequence-aware recommender) and spatial (e.g., POI recommender) data. Recommendation models are mainly categorized into collaborative filtering, content-based recommender system and hybrid recommender system based on the types of input data~\cite{adomavicius2005toward}.


Deep learning enjoys a massive hype at the moment. The past few decades have witnessed the tremendous success of the deep learning (DL) in many application domains such as computer vision and speech recognition. The academia and industry have been in a race to apply deep learning to a wider range of applications due to its capability in solving many complex tasks while providing start-of-the-art results~\cite{covington2016deep}. Recently, deep learning has been revolutionizing the recommendation architectures dramatically and brings more opportunities to improve the performance of recommender. Recent advances in deep learning based recommender systems have gained significant attention by overcoming obstacles of conventional models and achieving high recommendation quality. Deep learning is able to effectively capture the non-linear and non-trivial user-item relationships, and enable
the codification of more complex abstractions as data representations
in the higher layers. Furthermore, it catches the intricate relationships within the data itself, from abundant accessible data sources such as contextual, textual and visual information.

\paratitle{Pervasiveness and ubiquity of deep learning in recommender systems.} In industry, recommender systems are critical tools to enhance user experience and promote sales/services for many online websites and mobile applications~\cite{gomez2016netflix,Davidson:2010:YVR:1864708.1864770,covington2016deep,cheng2016wide,Shumpei2017}. For example, 80 percent of movies watched on Netflix came from recommendations~\cite{gomez2016netflix}, 60 percent of video clicks came from home page recommendation in YouTube~\cite{Davidson:2010:YVR:1864708.1864770}. Recently, many companies employ deep learning for further enhancing their recommendation quality~\cite{covington2016deep,cheng2016wide,Shumpei2017}. Covington et al.~\cite{covington2016deep} presented a deep neural network based recommendation algorithm for video recommendation on YouTube. Cheng et al.~\cite{cheng2016wide} proposed an App recommender system for Google Play with a wide \& deep model. Shumpei et al.~\cite{Shumpei2017} presented a RNN based news recommender system for Yahoo News. All of these models have stood the online testing and shown significant improvement over traditional models. Thus, we can see that deep learning has driven a remarkable revolution in industrial recommender applications.

The number of research publications on deep learning based recommendation methods has increased exponentially in these years, providing strong evidence of the inevitable pervasiveness of deep learning in recommender system research. The leading international conference on recommender system, RecSys\footnote{https://recsys.acm.org/}, started to organize regular workshop on deep learning for recommender system\footnote{http://dlrs-workshop.org/} since the year 2016. This workshop aims to promote research and encourage applications of deep learning based recommender system.

The success of deep learning for recommendation both in academia and in industry requires a comprehensive review and summary for successive researchers and practitioners to better understand the strength and weakness, and application scenarios of these models.

\paratitle{What are the differences between this survey and former ones?} Plenty of research has been done in the field of deep learning based recommendation. However, to the best of our knowledge, there are very few systematic reviews which well shape this area and position existing works and current progresses. Although some works have explored the recommender applications built on deep learning techniques and have attempted to formalize this research field, few has sought to provide an in-depth summary of current efforts or detail the open problems present in the area. This survey seeks to provide such a comprehensive summary of current research on deep learning based recommender systems, to identify open problems currently limiting real-world implementations and to point out future directions along this dimension.

In the last few years, a number of surveys in traditional recommender systems have been presented. For example, Su et al.~\cite{su2009survey} presented a systematic review on collaborative filtering techniques; Burke et al.~\cite{burke2002hybrid} proposed a comprehensive survey on hybrid recommender system; Fern{\'a}ndez-Tob{\'\i}as et al.~\cite{fernandez2012cross} and Khan et al.~\cite{Khan:2017:CDR:3101309.3073565} reviewed the cross-domain recommendation models; to name a few. However, there is a lack of extensive review on deep learning based recommender system. To the extent of our knowledge, only two related short surveys~\cite{Liu2017, 10.5120/ijca2017913361} are formally published. Betru et al.~\cite{10.5120/ijca2017913361} introduced three deep learning based recommendation models~\cite{salakhutdinov2007restricted, wang2015collaborative, van2013deep}, although these three works are influential in this research area, this survey lost sight of other emerging high quality works. Liu et al.~\cite{Liu2017} reviewed 13 papers on deep learning for recommendation, and proposed to classify these models based on the form of inputs (approaches using content information and approaches without content information) and outputs (rating and ranking). However, with the constant advent of novel research works, this classification framework is no longer suitable and a new inclusive framework is required for better understanding of this research field. Given the rising popularity and potential of deep learning applied in recommender system, a systematic survey will be of high scientific and practical values. We analyzed these works from different perspectives and presented some new insights toward this area. To this end, over 100 studies were shortlisted and classified in this survey.

\paratitle{How do we collect the papers?} In this survey, we collected over a hundred of related papers. We used Google Scholar as the main search engine, we also adopted the database, Web of Science, as an important tool to discover related papers. In addition, we screened most of the related high-profile conferences such as NIPS, ICML, ICLR, KDD, WWW, SIGIR, WSDM, RecSys, etc., just to name a few, to find out the recent work. The major keywords we used including: recommender system, recommendation, deep learning, neural networks, collaborative filtering, matrix factorization, etc.


\paratitle{Contributions of this survey.} The goal of this survey is to thoroughly review literature on the advances of deep learning based recommender system. It provides a panorama with which readers can quickly understand and step into the field of deep learning based recommendation. This survey lays the foundations to foster innovations in the area of recommender system and tap into the richness of this research area. This survey serves the researchers, practitioners, and educators who are interested in recommender system, with the hope that they will have a rough guideline when it comes to choosing the deep neural networks to solve recommendation tasks at hand. To summarize, the key contributions of this survey are three-folds: (1) We conduct a systematic review for  recommendation models based on deep learning techniques and propose a classification scheme to position and organize the current work; (2) We provide an overview and summary for the state-of-the-arts. (3) We discuss the challenges and open issues, and identify the new trends and future directions in this research field to share the vision and expand the horizons of deep learning based recommender system research.

The remaining of this article is organized as follows: Section 2 introduces the preliminaries for recommender systems and deep neural networks, we also discuss the advantages and disadvantages of deep neural network based recommendation models. Section 3 firstly presents our classification framework and then gives detailed introduction to the state-of-the-art. Section 4 discusses the challenges and prominent open research issues. Section 5 concludes the paper.


\section{Overview of Recommender Systems and Deep Learning}
Before we dive into the details of this survey, we start with an introduction to the basic terminology and concepts regarding recommender system and deep learning techniques. We also discuss the reasons and motivations of introducing deep neural networks to recommender systems.

\subsection{Recommender Systems}
Recommender systems estimate users' preference on items and recommend items that users might like to them proactively~\cite{adomavicius2005toward,ricci2015recommender}.  Recommendation models are usually classified into three categories~\cite{adomavicius2005toward,jannach2010recommender}: collaborative filtering, content based and hybrid recommender system. Collaborative filtering makes recommendations by learning from user-item historical interactions, either explicit (e.g. user's previous ratings) or implicit feedback (e.g. browsing history). Content-based recommendation is based primarily on comparisons across items' and users' auxiliary information. A diverse range of auxiliary information such as texts, images and videos can be taken into account. Hybrid model refers to recommender system that integrates two or more types of recommendation strategies~\cite{burke2002hybrid,jannach2010recommender}.

Suppose we have $M$ users and $N$ items, and $R$ denotes the interaction matrix and $\hat{R}$ denotes the predicted interaction matrix. Let $r_{ui}$ denote the preference of user $u$ to item $i$, and $\hat{r}_{ui}$ denote the predicted score. Meanwhile, we use a partially observed vector (rows of $R$) $\textbf{r}^{(u)} = \{r^{u1},...,r^{uN}\}$ to represent each user $u$, and partially observed vector (columns of $R$) $\textbf{r}^{(i)} = \{r^{1i},...,r^{Mi}\}$ to represent each item $i$.  $\mathcal{O}$ and $\mathcal{O^-}$ denote the observed and unobserved interaction set. we use $U \in \mathcal{R}^{M \times k}$ and $V \in \mathcal{R}^{N \times k}$ to denote user and item latent factor. $k$ is the dimension of latent factors. In addition, sequence information such as timestamp can also be considered to make sequence-aware recommendations. Other notations and denotations will be introduced in corresponding sections.

\subsection{Deep Learning Techniques}
Deep learning can be generally considered to be sub-field of machine learning. The typical defining essence of deep learning is that it learns \textit{deep representations}, i.e., learning multiple levels of representations and abstractions from data. For practical reasons, we consider any neural differentiable architecture as \textit{`deep learning`} as long as it optimizes a differentiable objective function using a variant of stochastic gradient descent (SGD). Neural architectures have demonstrated tremendous success in both supervised and unsupervised learning tasks~\cite{deng2014deep}. In this subsection, we clarify a diverse array of architectural paradigms that are closely related to this survey.
\begin{itemize}
    \item Multilayer Perceptron (MLP) is a feed-forward neural network with multiple (one or more) hidden layers between the input layer and output layer. Here, the perceptron can employ arbitrary activation function and does not necessarily represent strictly binary classifier. MLPs can be intrepreted as stacked layers of nonlinear transformations, learning hierarchical feature representations. MLPs are also known to be universal approximators.
    \item Autoencoder (AE) is an unsupervised model attempting to reconstruct its input data in the output layer. In general, the bottleneck layer (the middle-most layer) is used as a salient feature representation of the input data. There are many variants of autoencoders such as denoising autoencoder, marginalized denoising autoencoder, sparse autoencoder, contractive autoencoder and variational autoencoder (VAE)~\cite{Goodfellow-et-al-2016,chen2012marginalized}.
    \item Convolutional Neural Network (CNN)~\cite{Goodfellow-et-al-2016} is a special kind of feedforward neural network with convolution layers and pooling operations. It can capture the global and local features and significantly enhancing the efficiency and accuracy. It performs well in processing data with grid-like topology.
    \item Recurrent Neural Network (RNN)~\cite{Goodfellow-et-al-2016} is suitable for modelling sequential data. Unlike feedforward neural network, there are loops and memories in RNN to remember former computations. Variants such as Long Short Term Memory (LSTM) and Gated Recurrent Unit (GRU) network are often deployed in practice to overcome the vanishing gradient problem.
    \item Restricted Boltzmann Machine (RBM) is a two layer neural network consisting of a visible layer and a hidden layer. It can be easily stacked to a deep net. {\em Restricted} here means that there are no intra-layer communications in visible layer or hidden layer.
    \item Neural Autoregressive Distribution Estimation (NADE)~\cite{larochelle2011neural, uria2016neural} is an unsupervised neural network built atop autoregressive model and  feedforward neural networks. It is a tractable and efficient estimator for modelling data distribution and densities.
    \item Adversarial Networks (AN)~\cite{goodfellow2014generative} is a generative neural network which consists of a discriminator and a generator. The two neural networks are trained simultaneously by competing with each other in a minimax game framework.
    \item Attentional Models (AM) are differentiable neural architectures that operate based on soft content addressing over an input sequence (or image). Attention mechanism is typically ubiquitous and was incepted in Computer Vision and Natural Language Processing domains. However, it has also been an emerging trend in deep recommender system research.
    \item Deep Reinforcement Learning (DRL)~\cite{mnih2015human}. Reinforcement learning  operates on a trial-and-error paradigm. The whole framework mainly consists of the following components: agents, environments, states, actions and rewards. The combination between deep neural networks and reinforcement learning formulate DRL which have achieved human-level performance across multiple domains such as games and self-driving cars. Deep neural networks enable the agent to get knowledge from raw data and derive efficient representations without handcrafted features and domain heuristics.
\end{itemize}
Note that there are numerous advanced model emerging each year, here we only briefly listed some important ones. Readers who are interested in the details or more advanced models are referred to \cite{Goodfellow-et-al-2016}.

\subsection{Why Deep Neural Networks for Recommendation?}
Before diving into the details of recent advances, it is beneficial to understand the reasons of applying deep learning techniques to recommender systems. It is evident that numerous deep recommender systems have been proposed in a short span of several years. The field is indeed bustling with innovation. At this point, it would be easy to question the \textit{need} for so many different architectures and/or possibly even the utility of neural networks for the problem domain. Along the same tangent, it would be apt to provide a clear rationale of why each proposed architecture and to which scenario it would be most beneficial for. All in all, this question is highly relevant to the issue of task, domains and recommender scenarios.
One of the most attractive properties of neural architectures is that they are (1) end-to-end differentiable and (2) provide suitable \textit{inductive biases} catered to the input data type. As such, if there is an inherent structure that the model can exploit, then deep neural networks ought to be useful. For instance, CNNs and RNNs have long exploited the instrinsic structure in vision (and/or human language). Similarly, the sequential structure of session or click-logs are highly suitable for the inductive biases provided by recurrent/convolutional models \cite{tang2018personalized,hidasi2015session,wu2017recurrent}.

Moreover, deep neural networks are also composite in the sense that multiple neural building blocks can be composed into a single (gigantic) differentiable function and trained end-to-end. The key advantage here is when dealing with \textit{content-based} recommendation. This is inevitable when modeling users/items on the web, where multi-modal data is commonplace. For instance, when dealing with textual data (reviews \cite{Zheng:2017:JDM:3018661.3018665}, tweets \cite{gong2016hashtag} etc.), image data (social posts, product images), CNNs/RNNs become indispensable neural building blocks. Here, the traditional alternative (designing modality-specific features etc.) becomes significantly less attractive and consequently, the recommender system cannot take advantage of joint (end-to-end) representation learning. In some sense, developments in the field of recommender systems are also tightly coupled with advances research in related modalities (such as vision or language communities). For example, to process reviews, one would have to perform costly preprocessing (e.g., keyphrase extraction, topic modeling etc.) whilst newer deep learning-based approaches are able to ingest all textual information end-to-end \cite{Zheng:2017:JDM:3018661.3018665}. All in all, the capabilities of deep learning in this aspect can be regarded as paradigm-shifting and the ability to represent images, text and interactions in a unified joint framework \cite{zhang2017joint} is not possible without these recent advances.

Pertaining to the interaction-only setting (i.e., matrix completion or collaborative ranking problem), the key idea here is that deep neural networks are justified when there is a huge amount of complexity or when there is a large number of training instances. In \cite{he2017neural}, the authors used a MLP to approximate the interaction function and showed reasonable performance gains over traditional methods such as MF. While these neural models perform better, we also note that standard machine learning models such as BPR, MF and CML are known to perform reasonably well when trained with momentum-based gradient descent on interaction-only data \cite{Tay:2018:LRM:3178876.3186154}. However, we can also consider these models to be also neural architectures as well, since they take advantage of recent deep learning advances such as Adam, Dropout or Batch Normalization \cite{he2017neural,zhang2018neurec}. It is also easy to see that, traditional recommender algorithms (matrix factorization, factorization machines, etc.) can also be expressed as neural/differentiable architectures \cite{he2017neural,He2017fmneural} and trained efficiently with a framework such as Tensorflow or Pytorch, enabling efficient GPU-emabled training and free automatic differentiation. Hence, in today's research climate (and even industrial), there is completely \textit{no reason} to not used deep learning based tools for development of any recommender system.

To recapitulate,  we summarize the strengths of deep learning based recommendation models that readers might bear in mind when try to employ them for practice use.
\raggedbottom
\begin{itemize}
    \item \textbf{Nonlinear Transformation}. Contrary to linear models, deep neural networks is capable of modelling the non-linearity in data with nonlinear activations  such as relu, sigmoid, tanh, etc. This property makes it possible to capture the complex and intricate user item interaction patterns. Conventional methods such as matrix factorization, factorization machine, sparse linear model are essentially linear models. For example, matrix factorization models the user-item interaction by linearly combining user and item latent factors~\cite{he2017neural}; Factorization machine is a member of multivariate linear family~\cite{He2017fmneural}; Obviously, SLIM is a linear regression model with sparsity constraints. The linear assumption, acting as the basis of many traditional recommenders, is oversimplified and will greatly limit their modelling expressiveness. It is well-established that neural networks are able to approximate any continuous function with an arbitrary precision by varying the activation choices and combinations~\cite{hornik1989multilayer,hornik1991approximation}. This property makes it possible to deal with complex interaction patterns and precisely reflect user's preference.

    \item  \textbf{Representation Learning}. Deep neural networks is efficacious in learning the underlying explanatory factors and useful representations from input data. In general,  a large amount of descriptive information about items and users is available in real-world applications. Making use of this information provides a way to advance our understanding of items and users, thus, resulting in a better recommender. As such, it is a natural choice to apply deep neural networks to representation learning in recommendation models. The advantages of using deep neural networks to assist representation learning are in two-folds: (1) it reduces the efforts in hand-craft feature design.  Feature engineering is a labor intensive work, deep neural networks enable automatically feature learning from raw data in unsupervised or supervised approach; (2) it enables recommendation models to include heterogeneous content information such as text, images, audio and even video. Deep learning networks have made breakthroughs in multimedia data processing and shown potentials in representations learning from various sources.

    \item  \textbf{Sequence Modelling}. Deep neural networks have shown promising results on a number of sequential modelling tasks such as machine translation, natural language understanding, speech recognition, chatbots, and many others.  RNN and CNN play critical roles in these tasks.  RNN achives this with internal memory states while CNN achieves this with filters sliding along with time. Both of them are widely applicable and flexible in mining sequential structure in data.   Modelling sequential signals is an important topic for mining the temporal dynamics of user behaviour and item evolution. For example, next-item/basket prediction and session based recommendation are typical applications. As such, deep neural networks become a perfect fit for this sequential pattern mining task. This

    \item \textbf{Flexibility}. Deep learning techniques possess high flexibility, especially with the advent of many popular deep learning frameworks such as Tensorflow\footnote{https://www.tensorflow.org/}, Keras\footnote{https://keras.io/}, Caffe\footnote{http://caffe.berkeleyvision.org/}, MXnet\footnote{https://mxnet.apache.org/}, DeepLearning4j\footnote{https://deeplearning4j.org/}, PyTorch\footnote{https://pytorch.org/}, Theano\footnote{http://deeplearning.net/software/theano/}, etc.  Most of these tools are developed in a modular way and have active community and professional support. The good modularization makes development and engineering a lot more efficient. For example, it is easy to combine different neural structures to formulate powerful hybrid models, or replace one module with others. Thus, we could easily build hybrid and composite recommendation models to simultaneously capture different characteristics and factors.
\end{itemize}
\subsection{On Potential Limitations}
 Are there really any drawbacks and limitations with using deep learning for recommendation? In this section, we aim to tackle several commonly cited arguments against the usage of deep learning for recommender systems research.
\begin{itemize}
    \item \textbf{Interpretability.} Despite its success, deep learning is well-known to behave as black boxes, and providing explainable predictions seem to be a really challenging task. A common argument against deep neural networks is that the hidden weights and activations are generally non-interpretable, limiting explainability. However, this concern has generally been eased with the advent of neural attention models and have paved the world for deep neural models that enjoy improved interpretability \cite{seo2017interpretable,xiao2017attentional,Tay:2018:MCN:3219819.3220086}. While interpreting individual neurons still pose a challenge for neural models (not only in recommender systems), present state-of-the-art models are already capable of some extent of interpretability, enabling explainable recommendation. We discuss this issue in more detail in the open issues section.

    \item \textbf{Data Requirement.}
    A second possible limitation is that deep learning is known to be data-hungry, in the sense that it requires sufficient data in order to fully support its rich parameterization. However, as compared with other domains (such as language or vision) in which labeled data is scarce, it is relatively easy to garner a significant amount of data within the context of recommender systems research. Million/billion scale datasets are commonplace not only in industry but also released as academic datasets.

    \item \textbf{Extensive Hyperparameter Tuning.}
    A third well-established argument against deep learning is the need for extensive hyperparameter tuning. However, we note that hyperparameter tuning is not an exclusive problem of deep learning but machine learning in general (e.g., regularization factors and learning rate similarly have to be tuned for traditional matrix factorization etc)  Granted, deep learning may introduce additional hyperparameters in some cases. For example, a recent work \cite{Tay:2018:LRM:3178876.3186154}, attentive extension of the traditional metric learning algorithm \cite{hsieh2017collaborative} only introduces a single hyperparameter.

\end{itemize}

\section{Deep Learning Based Recommendation: State-of-the-art}
In this section, we we firstly introduce the categories of deep learning based recommendation models and then highlight state-of-the-art research prototypes, aiming to identify the most notable and promising advancement in recent years.

\subsection{Categories of deep learning based recommendation models}
\begin{figure*}
\includegraphics[width=0.80\textwidth]{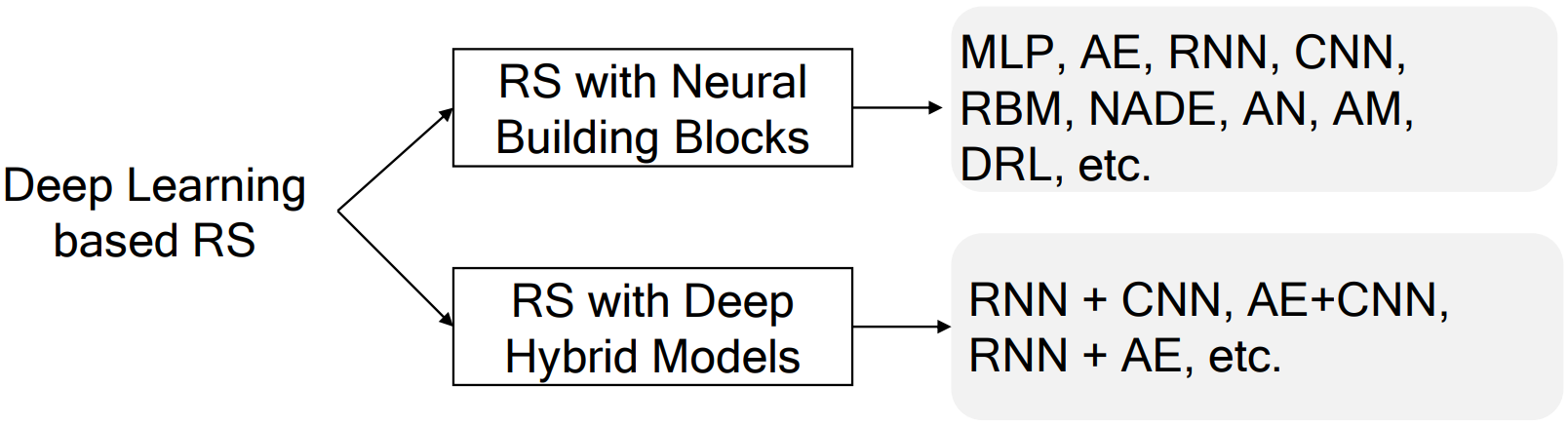}
\caption{Categories of deep neural network based recommendation models.}
\label{classification_sch}
\end{figure*}
To provide a bird-eye's view of this field, we classify the existing models based the types of employed deep learning techniques. We further divide deep learning based recommendation models into the following two categories. Figure \ref{classification_sch} summarizes the classification scheme.

\begin{table}[]
\centering
\caption{A lookup table for reviewed publications.}
\label{categories}
\begin{tabular}{|c|l|}
\hline
Categories & Publications \\ \hline
MLP     &   \begin{tabular}[l]{@{}l@{}} \cite{huang2015neural,carl2017bridging,alashkar2017examples,Vuurens:2016:EDS:2988450.2988457,wang2017item,he2017neural,Chen:2017:LCD:3041021.3054227,ebesu2017neural,cheng2016wide,lian2017cccfnet,covington2016deep,guo2017deepfm,liang2015content,He2017fmneural},\\\cite{elkahky2015multi,song2018neural,chen2017location,xu2016tag,xu2017tag,niu2018neural,lian2018xdeepfm,vartak2017meta} \end{tabular}      \\ \hline
Autoencoder      &    \begin{tabular}[l]{@{}l@{}}    \cite{ouyang2014autoencoder,sedhain2015autorec,strub2015collaborative,wu2016collaborative,strub2016hybrid,yi2016expanded,zhuang2017representationRanking,suzuki2017stacked,pana2017trust,wang2015collaborative,li2016cvae,li2015deep,dong2017hybrid},\\\cite{zhuang2017representation,wang2015relational,bai2017dltsr,ying2016collaborative,zhang2017autosvd++,wei2016collaborative,wei2017collaborative,unger2015latent,cao2017online,zuo2016tag,unger2016towards,deng2017deep,liang2018variational}                     \end{tabular}           \\ \hline
CNNs      &    \begin{tabular}[l]{@{}l@{}}
\cite{kim2016convolutional,kim2017deep,Zheng:2017:JDM:3018661.3018665,van2013deep,seo2017representation,deepstyle,chu2017hybrid,shen2016automatic,zhou2016applying,mcauley2015image,he2016ups,he2016vbpr,wen2016visual,Wang:2017:YIR:3038912.3052638},\\\cite{gong2016hashtag,Nguyen2017,wang2017dynamicattention,yu2018aesthetic,he2018outer,tang2018personalized,tuan20173d,ying2018graph,berg2017graph,lee2018collaborative,seo2017interpretable}
\end{tabular}         \\ \hline
RNNs                   &    \begin{tabular}[l]{@{}l@{}}\cite{bansal2016ask,ko2016collaborative,smirnova2017contextual,dai2016deep,li2016hashtag,tan2016improved,wu2016joint,jing2017neural,hidasi2016parallel,wu2016personal,wu2017recurrent,hidasi2015session,suglia2017deep,quadrana2017personalizing,donkers2017sequential},\\\cite{tan2016neural,musto2016ask,xie2016neural,wu2016recurrent,devooght2016collaborative,soh2017deep,dai2016recurrent,twardowski2016modelling,Shumpei2017,hidasi2017recurrent,Jannach:2017:RNN:3109859.3109872,li2018learning,christakopoulou2018q}
\end{tabular}         \\ \hline
RBM                   &    \begin{tabular}[c]{@{}c@{}}\cite{salakhutdinov2007restricted,georgiev2013non,liu2015item,xie2016user,jia2016collaborative,wang2014improving,jia2015multi}                          \end{tabular}           \\ \hline
NADE                  &   \cite{Zheng:2016:NAA:3045390.3045472,Zheng:2016:NAC:2988450.2988453,du2016collaborative}          \\ \hline
Neural Attention                   &\begin{tabular}[l]{@{}l@{}} \cite{chen2017acf,Tay:2018:LRM:3178876.3186154,jhamb2018attentive,gong2016hashtag,seo2017representation,wang2017dynamicattention,li2016hashtag,loyola2017modeling,liu2018stamp,ying2018sequential,zhou2017atrank,zhang2018dynamic},\\\cite{Tay:2018:MCN:3219819.3220086,zhanghashtag,hu2018leveraging}  \end{tabular} \\ \hline
Adversary Network                   &     \cite{wang2017irgan,he2018adversarial,Cai2018GenerativeAN,Wang:2018:NMS:3219819.3220004}       \\ \hline
DRL                   &     \cite{zhao2018deep,zhao2018recommendations,zheng2018drn,munemasadeep,choi2018reinforcement,wang2014exploration,chen2018stabilizing}          \\ \hline
Hybrid Models         & \begin{tabular}[c]{@{}c@{}}\cite{lei2016comparative,rawat2016contagnet,zhanghashtag,song2016multi,zhang2016collaborative,ebesu2017neural,li2017neural,lee2016quote,gao2014modeling,wang2016collaborative,Chen:2017:PKF:3077136.3080776}   \end{tabular}     \\ \hline
\end{tabular}
\end{table}

\begin{itemize}
    \item \textit{Recommendation with Neural Building Blocks}. In this category, models are divided into eight subcategories in conformity with the aforementioned eight deep learning models: MLP, AE, CNNs, RNNs, RBM, NADE, AM, AN and DRL based recommender system. The deep learning technique in use determines the applicability of recommendation model. For instance, MLP can easily model the non-linear interactions between users and items; CNNs are capable of extracting local and global representations from heterogeneous data sources such as textual and visual information; RNNs enable the recommender system to model the temporal dynamics and sequential evolution of content information.

    \item \textit{Recommendation with Deep Hybrid Models}. Some deep learning based recommendation models utilize more than one deep learning technique. The flexibility of deep neural networks makes it possible to combine several neural building blocks together to complement one another and form a more powerful hybrid model. There are many possible combinations of these night deep learning techniques but not all have been exploited. Note that it is different from the hybrid deep networks in \cite{deng2014deep} which refer to the deep architectures that make use of both generative and discriminative components.

\end{itemize}

\begin{table}[]
\centering
\caption{Deep neural network based recommendation models in specific application fields.}
\label{applicationfields}
\begin{tabular}{|c|c|l|}
\hline
   \begin{tabular}[c]{@{}c@{}}Data \\Sources/Tasks\end{tabular}                                                                   & Notes                                                               & \multicolumn{1}{l|}{Publications} \\ \hline
\multirow{3}{*}{\begin{tabular}[c]{@{}c@{}}Sequential\\ Information\end{tabular}} & w/t User ID                                                         & \multicolumn{1}{l|}{\cite{zhang2018dynamic,quadrana2017personalizing,jing2017neural,tang2018personalized,ying2018sequential,donkers2017sequential,soh2017deep,wu2016recurrent,wu2017recurrent,devooght2016collaborative,dai2016recurrent,li2018learning,zhou2017atrank,zhao2018deep,chen2018stabilizing,wang2016collaborative}}             \\ \cline{2-3}   & \begin{tabular}[c]{@{}c@{}}Session based\\ w/o User ID\end{tabular} &\cite{hidasi2015session,tan2016improved,twardowski2016modelling,hidasi2016parallel,jing2017neural,hidasi2017recurrent,quadrana2017personalizing,Loyola:2017:MUS:3109859.3109917,loyola2017modeling,liu2018stamp,Jannach:2017:RNN:3109859.3109872,tuan20173d}           \\ \cline{2-3}
 & Check-In, POI & \cite{carl2017bridging,Wang:2017:YIR:3038912.3052638,unger2015latent,unger2016towards}    \\ \cline{2-3}
                                                                                  \hline
\multirow{5}{*}{Text}                                            & Hash Tags   & \cite{zhanghashtag,rawat2016contagnet,gong2016hashtag,wang2015relational,xu2017tag,xu2016tag,zuo2016tag,Nguyen2017} \\ \cline{2-3}
& News   & \cite{chen2017location,wang2017dynamicattention,Shumpei2017,song2016multi,zheng2018drn,cao2017online}  \\ \cline{2-3}
& Review texts  &\cite{Zheng:2017:JDM:3018661.3018665,catherine2017transnets,Tay:2018:MCN:3219819.3220086,zhang2017joint,li2017neural,wu2016joint,seo2017interpretable} \\\cline{2-3}
& Quotes & \cite{lee2016quote,tan2016neural} \\\cline{2-3} \hline
Images    &    Visual features       &     \cite{zhang2017joint,lei2016comparative,wen2016visual,xie2016neural,yu2018aesthetic,niu2018neural,Wang:2017:YIR:3038912.3052638,mcauley2015image,he2016ups,alashkar2017examples,deepstyle,chu2017hybrid,zhou2016applying,he2016vbpr,chen2017acf,zhang2016collaborative}                                                                                                 \\ \hline
Audio    &  Music  &    \cite{van2013deep,liang2015content,wang2014improving,wang2014exploration}                               \\ \hline
Video    & Videos   &      \cite{Chen:2017:PKF:3077136.3080776,lee2018collaborative,covington2016deep,chen2017acf}     \\ \hline
\multirow{3}{*}{Networks}    & Citation Network  &    \cite{ebesu2017neural,huang2015neural,Cai2018GenerativeAN}       \\\cline{2-3}
& Social Network   &    \cite{wang2017item,deng2017deep,pana2017trust}     \\\cline{2-3}
& Cross Domain   &    \cite{lian2017cccfnet,elkahky2015multi,wang2017item}     \\\cline{2-3}
\hline
\multirow{3}{*}{Others}    & Cold-start   &     \cite{vartak2017meta,volkovs2017dropoutnet,wei2016collaborative,wei2017collaborative}   \\\cline{2-3}
& Multitask   &    \cite{bansal2016ask,li2017neural,jing2017neural,wu2016joint,yi2016expanded}     \\\cline{2-3}
& Explainability   &  \cite{li2017neural,seo2017interpretable}      \\\cline{2-3}
\hline
\end{tabular}
\end{table}

Table \ref{categories} lists all the reviewed models, we organize them following the aforementioned classification scheme. Additionally, we also summarize some of the publications from the task perspective in Table \ref{applicationfields}. The reviewed publications are concerned with a variety of tasks. Some of the tasks have started to gain attention due to use of deep neural networks such as session-based recommendation, image, video recommendations. Some of the tasks might not be novel to the recommendation research area (a detail review on the side information for recommender systems can be found in \cite{shi2014collaborative} ), but DL provides more possibility to find better solutions. For example, dealing with images and videos would be tough task without the help of deep learning techniques. The sequence modelling capability of deep neural networks makes it easy to capture the sequential patterns of user behaviors. Some of the specific tasks will be discussed in the following text.


\subsection{Multilayer Perceptron based Recommendation}
MLP is a concise but effective network which has been demonstrated to be able to approximate any measurable function to any desired degree of accuracy~\cite{hornik1989multilayer}. As such, it is the basis of numerous advanced approaches and is widely used in many areas.

\paratitle{Neural Extension of Traditional Recommendation Methods}. Many existing recommendation models are essentially linear methods. MLP can be used to add nonlinear transformation to existing RS approaches and interpret them into neural extensions.


\begin{figure}
\begin{center}
\begin{minipage}[t]{5.7cm}
\includegraphics[width=5.7cm]{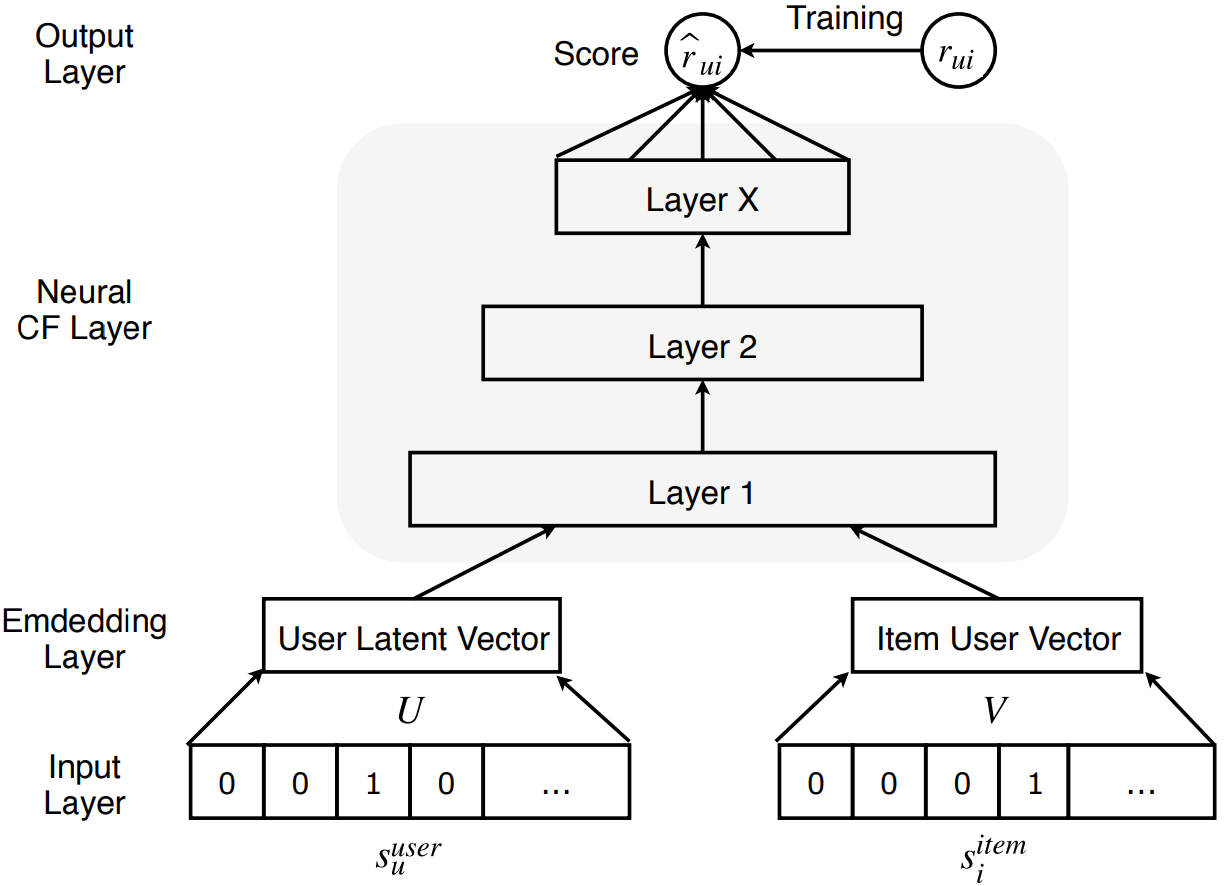}
\centering{(a)}
\end{minipage}
\hspace{1cm}
\begin{minipage}[t]{6.0cm}
\includegraphics[width=6.0cm]{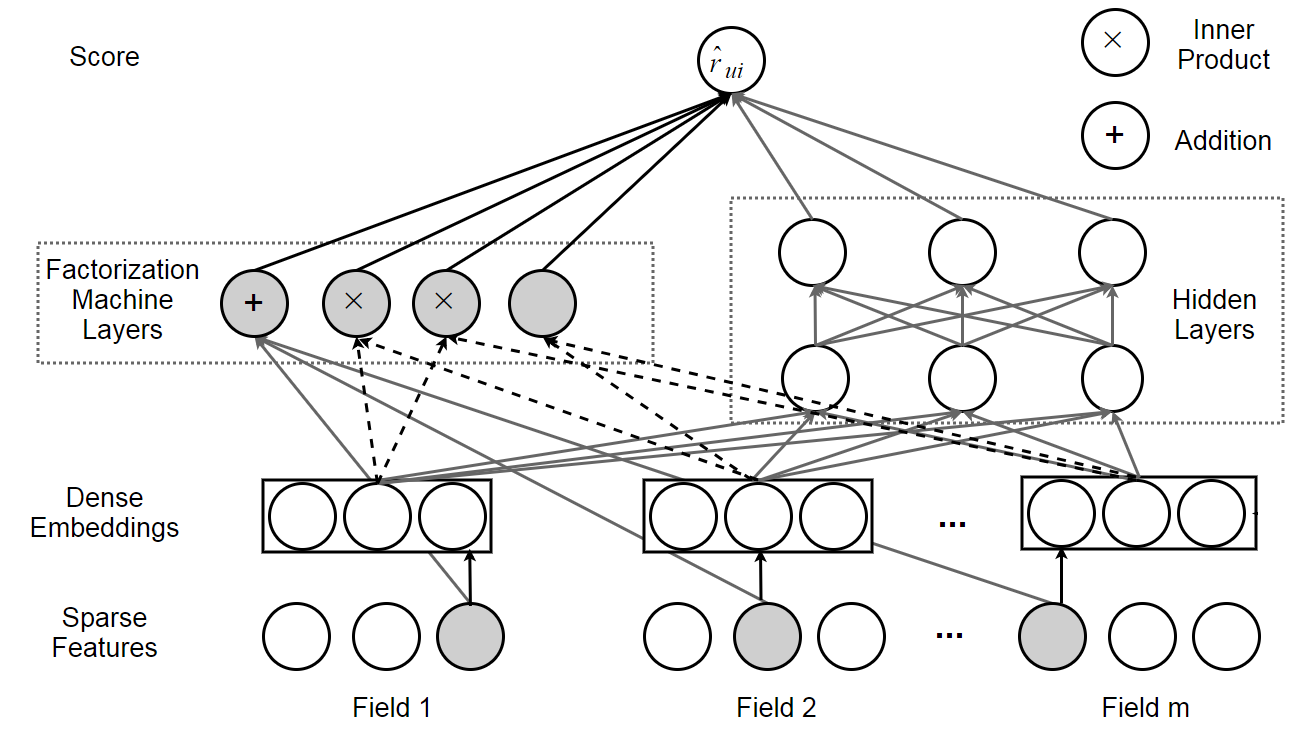}
\centering{(b)}
\end{minipage}
\caption{ Illustration of: (a) Neural Collaborative Filtering; (b) Deep Factorization Machine.}
\label{fig:mlp}
\end{center}
\end{figure}

\textit{Neural Collaborative Filtering}. In most cases, recommendation is deemed to be a two-way interaction between users preferences and items features. For example, matrix factorization decomposes the rating matrix into low-dimensional user/item latent factors.  It is natural to construct a dual neural network to model the two-way interaction between users and items. Neural Network Matrix Factorization (NNMF)~\cite{dziugaite2015neural} and Neural Collaborative Filtering (NCF)~\cite{he2017neural} are two representative works. Figure \ref{fig:mlp}a shows the NCF architecture.
Let $s_u^{user}$ and $s_i^{item}$ denote the side information (e.g. user profiles and item features), or just one-hot identifier of user $u$ and item $i$. The scoring function is defined as follows:
\begin{equation}
\hat{r}_{ui} = f(U^T \cdot s_u^{user}, V^T \cdot s_i^{item}|U,V,\theta)
\end{equation}
where function $f(\cdot)$ represents the multilayer perceptron, and $\theta$ is the parameters of this network. Traditional MF can be viewed as a special case of NCF. Therefore, it is convenient to fuse the neural interpretation of matrix factorization with MLP to formulate a more general model which makes use of both linearity of MF and non-linearity of MLP to enhance recommendation quality. The whole network can be trained with weighted square loss (for explicit feedback) or binary cross-entropy loss (for implicit feedback). The cross-entropy loss is defined as:
\begin{equation}
\mathcal{L} = -\sum_{(u,i) \in \mathcal{O} \cup \mathcal{O^-}} r_{ui}\log\hat{r}_{ui} + (1-r_{ui})\log(1-\hat{r}_{ui})
\end{equation}
Negative sampling approaches can be used to reduce the number of training unobserved instances. Follow-up work~\cite{niu2018neural,song2018neural} proposed using pairwise ranking loss to enhance the performance. He et al.~\cite{wang2017item,lian2017cccfnet} extended the NCF model to cross-domain recommendations. Xue et al.~\cite{xue2017deep} and Zhang et al.~\cite{zhang2018neurec} showed that the one-hot identifier can be replaced with columns or rows of the interaction matrix to retain the user-item interaction patterns.


\textit{Deep Factorization Machine}. DeepFM~\cite{guo2017deepfm} is an end-to-end model which seamlessly integrates factorization machine and MLP. It is able to model the high-order feature interactions via deep neural network and low-order interactions with factorization machine. Factorization machine (FM) utilizes addition and inner product operations to capture the linear and pairwise interactions between features (refer to Equation (1) in \cite{rendle2010factorization} for more details). MLP leverages the non-linear activations and deep structure to model the high-order interactions. The way of combining MLP with FM is enlightened by wide \& deep network.  It replaces the wide component with a neural interpretation of factorization machine. Compared to wide \& deep model, DeepFM does not require tedious feature engineering. Figure \ref{fig:mlp}b illustrates the structure of DeepFM.  The input of DeepFM $x$ is an $m$-fields data consisting of pairs $(u, i)$ (identity and features of user and item). For simplicity, the outputs of FM and MLP are denoted as $y_{FM}(x)$ and $y_{MLP}(x)$ respectively. The prediction score is calculated by:
\begin{equation}
\hat{r}_{ui} = \sigma(y_{FM}(x) + y_{MLP}(x))
\end{equation}
where $\sigma(\cdot)$ is the sigmoid activation function.

Lian et al.~\cite{lian2018xdeepfm} improved DeepMF by proposing a eXtreme deep factorization machine to jointly model the explicit and implicit feature interactions. The explicit high-order feature interactions are learned via a compressed interaction network. A parallel work proposed by He et al.~\cite{He2017fmneural} replaces the second-order interactions with MLP and proposed regularizing the model with dropout and batch normalization.

\paratitle{Feature Representation Learning with MLP}. Using MLP for feature representation is very straightforward and highly efficient, even though it might not be as expressive as autoencoder, CNNs and RNNs.

\begin{figure}
\begin{center}
\begin{minipage}[t]{5.9cm}
\includegraphics[width=5.9cm]{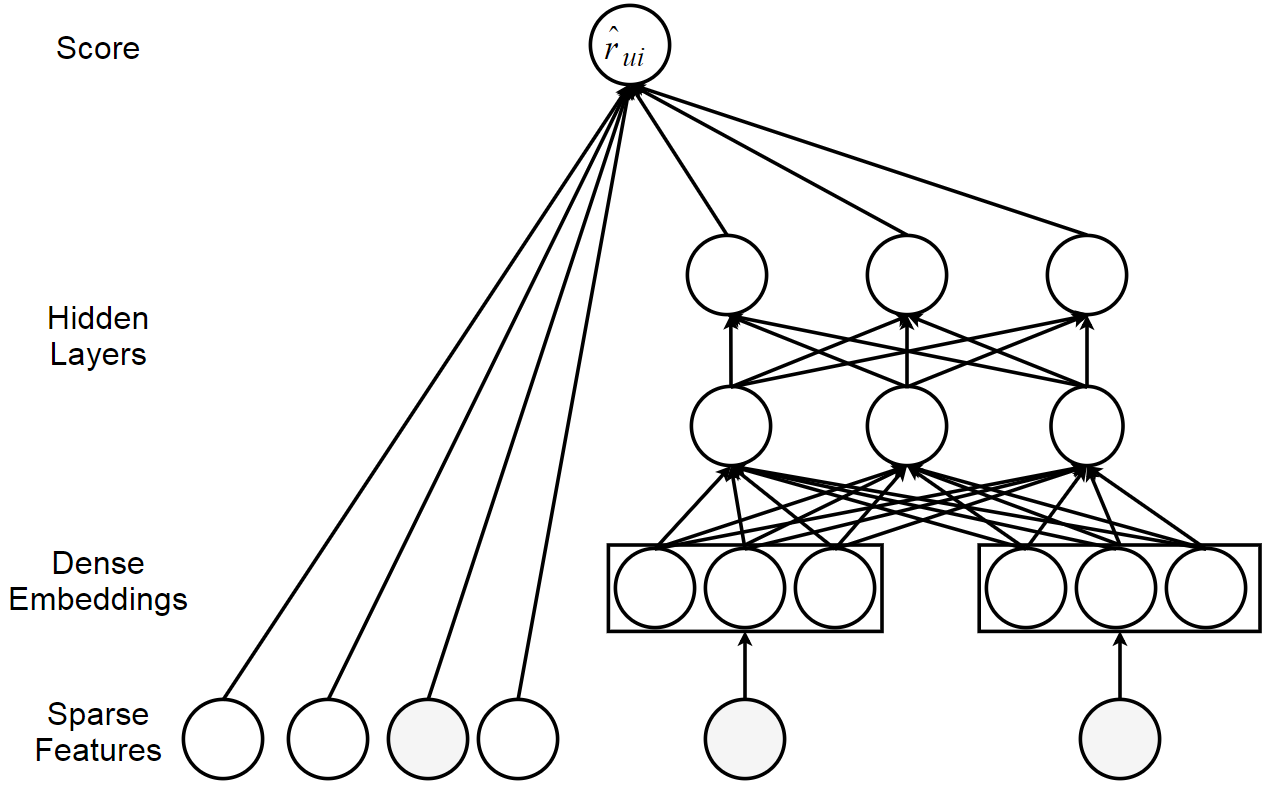}
\centering{(a)}
\end{minipage}
\hspace{1cm}
\begin{minipage}[t]{6.0cm}
\includegraphics[width=6.0cm]{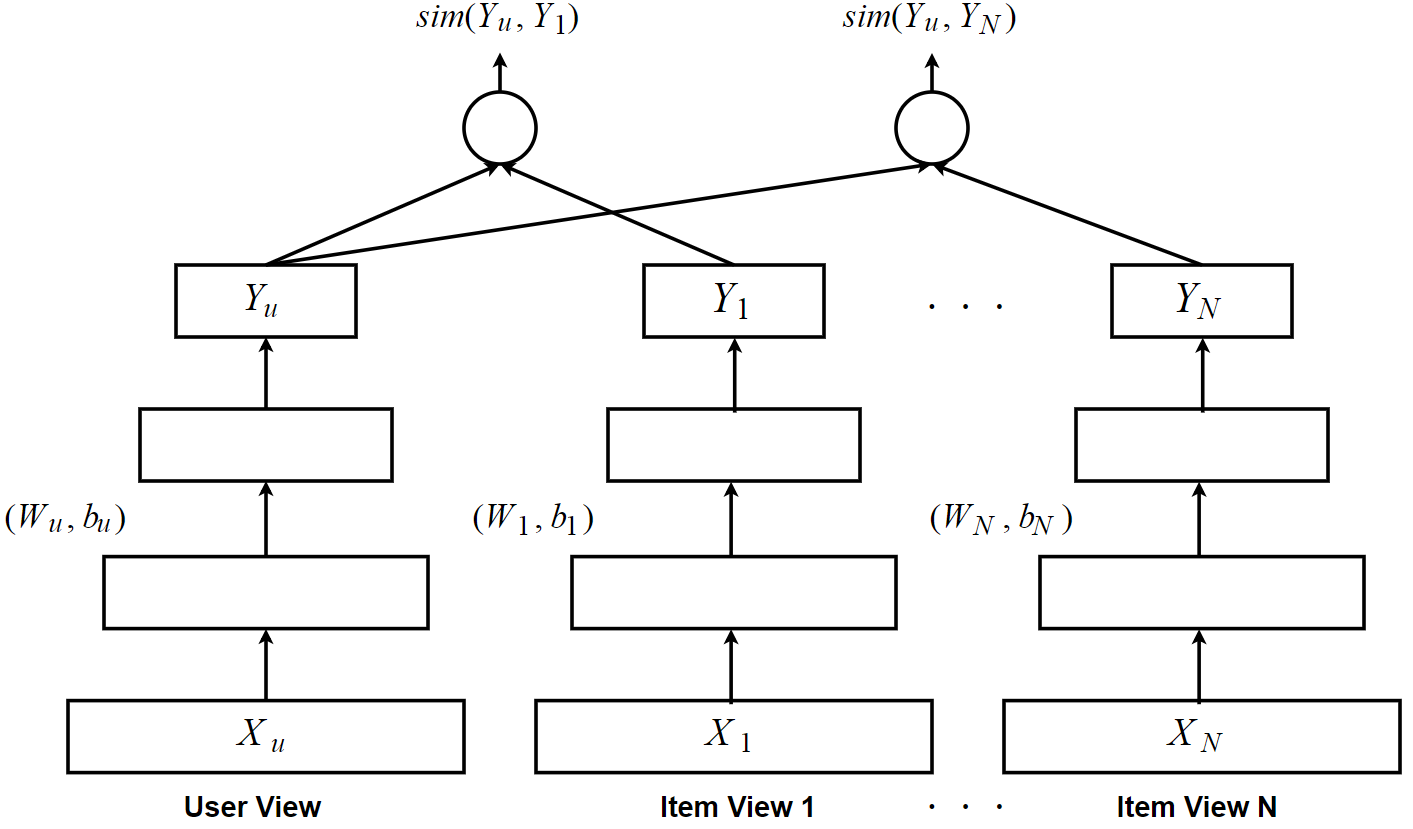}
\centering{(b)}
\end{minipage}
\caption{ Illustration of: (a) Wide \& Deep Learning; (b) Multi-View Deep Neural Network.}
\label{fig:mlp2}
\end{center}
\end{figure}

\textit{Wide \& Deep Learning}. This general model (shown in Figure \ref{fig:mlp2}a) can solve both regression and classification problems, but initially introduced for App recommendation in Google play~\cite{cheng2016wide}. The wide learning component is a single layer perceptron which can also be regarded as a generalized linear model. The deep learning component is multilayer perceptron. The rationale of combining these two learning techniques is that it enables the recommender to capture both memorization and generalization. Memorization achieved by the wide learning component represents the capability of catching the direct features from historical data. Meanwhile, the deep learning component catches the generalization by producing more general and abstract representations. This model can improve the accuracy as well as the diversity of recommendation.

Formally, the wide learning is defined as: $y = W^T_{wide}\{x,\phi(x)\} +b$, where $W^T_{wide}$, $b$ are the model parameters. The input $\{x,\phi(x)\}$ is the concatenated feature set consisting of raw input feature $x$ and transformed (e.g. cross-product transformation to capture the correlations between features) feature $\phi(x)$. Each layer of the deep neural component is in the form of $\alpha^{(l + 1)} =f(W^{(l)}_{deep}a^{(l)} +b^{(l)})$, where $l$ indicates the $l^{th}$ layer, and $f(\cdot)$ is the activation function. $W^{(l)}_{deep}$ and $b^{(l)}$ are weight and bias terms.  The wide \& deep learning model is attained by fusing these two models:
\begin{equation}
P(\hat{r}_{ui}=1|x) = \sigma(W^T_{wide}\{x,\phi(x)\} + W^T_{deep} a^{(l_f)}+ bias)
\end{equation}
where $\sigma(\cdot)$ is the sigmoid function, $\hat{r}_{ui}$ is the binary rating label, $a^{(l_f)}$ is the final activation. This joint model is optimized with stochastic back-propagation ( follow-the-regularized-leader algorithm). Recommending list is generated based on the predicted scores.

By extending this model, Chen et al.~\cite{Chen:2017:LCD:3041021.3054227} devised a locally-connected wide \& deep learning model for large scale industrial-level recommendation task. It employs the efficient locally-connected network to replace the deep learning component, which decreases the running time by one order of magnitude. An important step of deploying wide \& deep learning is selecting features for wide and deep parts. In other word, the system should be able to determine which features are memorized or generalized. Moreover, the cross-product transformation also is required to be manually designed. These pre-steps will greatly influence the utility of this model. The above mentioned deep factorization based model can alleviate the effort in feature engineering.

Covington et al.~\cite{covington2016deep} explored applying MLP in YouTube recommendation. This system divides the recommendation task into two stages: candidate generation and candidate ranking. The candidate generation network retrieves a subset (hundreds) from all video corpus. The ranking network generates a top-n list (dozens) based on the nearest neighbors scores from the candidates. We notice that the industrial world cares more about feature engineering (e.g. transformation, normalization, crossing) and scalability of recommendation models.

Alashkar et al.~\cite{alashkar2017examples} proposed a MLP based model for makeup recommendation. This work uses two identical MLPs to model labeled examples and expert rules respectively. Parameters of these two networks are updated simultaneously by minimizing the differences between their outputs. It demonstrates the efficacy of adopting expert knowledge to guide the learning process of the recommendation model in a MLP framework. It is highly precise even though the expertise acquisition needs a lot of human involvements.

\textit{Collaborative Metric Learning (CML)}. CML~\cite{hsieh2017collaborative} replaces the dot product of MF with Euclidean distance because dot product does not satisfy the triangle inequality of distance function. The user and item embeddings are learned via maximizing the distance between users and their disliked items and minimizing that between users and their preferred items. In CML, MLP is used to learn representations from item features such as text, images and tags.

\paratitle{Recommendation with Deep Structured Semantic Model}.
Deep Structured Semantic Model
(DSSM)~\cite{huang2013learning} is a deep neural network for learning semantic representations of entities in a common continuous semantic space and measuring their semantic similarities. It is widely used in information retrieval area and is supremely suitable for top-n recommendation~\cite{ elkahky2015multi, xu2016tag}. DSSM projects different entities into a common low-dimensional space, and computes their similarities with cosine function. Basic DSSM is made up of MLP so we put it in this section. Note that, more advanced neural layers such as convolution and max-pooling layers can also be easily integrated into DSSM.

\textit{Deep Semantic Similarity based Personalized Recommendation (DSPR)}~\cite{xu2016tag} is a tag-aware personalized recommender where each user $x_u$ and item $x_i$ are represented by tag annotations and mapped into a common tag space. Cosine similarity $sim(u,i)$ are applied to decide the relevance of items and users (or user's preference over the item). The loss function of DSPR is defined as follows:
\begin{equation}
\mathcal{L} = - \sum_{(u,i*)}[log(e^{sim(u,i*)}) - log(\sum_{(u,i^-)\in D^-} e^{sim(u,i^-)} )]
\end{equation}
where $(u,i^-)$ are negative samples which are randomly sampled from the negative user item pairs. The authors.~\cite{xu2017tag} further improved DSPR using autoencoder to learn low-dimensional representations from user/item profiles.

\textit{Multi-View Deep Neural Network (MV-DNN)}~\cite{elkahky2015multi} is designed for cross domain recommendation. It treats users as the pivot view and each domain (suppose we have $Z$ domains) as auxiliary view. Apparently, there are $Z$ similarity scores for $Z$ user-domain pairs. Figure \ref{fig:mlp2}b illustrates the structure of MV-DNN. The loss function of MV-DNN is defined as:
\begin{equation}
\mathcal{L} = \underset{\theta}{arg min}\sum_{j=1}^Z \frac{ exp(\gamma \cdot cosine(Y_u, Y_{a,j}))}{\sum_{X' \in R^{da}}exp(\gamma \cdot cosine(Y_u, f_a(X')))}
\end{equation}
where $\theta$ is the model parameters, $\gamma$ is the smoothing factor, $Y_u$ is the output of user view,  $a$ is the index of active view. $R^{da}$ is the input domain of view $a$. MV-DNN is capable of scaling up to many domains. However, it is based on the hypothesis that users have similar tastes in one domain should have similar tastes in other domains. Intuitively, this assumption might be unreasonable in many cases. Therefore, we should have some preliminary knowledge on the correlations across different domains to make the most of MV-DNN.

\subsection{Autoencoder based Recommendation}
There exist two general ways of applying autoencoder to recommender system: (1) using autoencoder to learn lower-dimensional feature representations at the bottleneck layer; or (2) filling the blanks of the interaction matrix directly in the reconstruction layer. Almost all the autoencoder variants such as denoising autoencoder, variational autoencoder, contactive autoencoder and marginalized autoencoder can be applied to recommendation task. Table \ref{table:autoencoder} summarizes the recommendation models based on the types of autoencoder in use.


\paratitle{Autoencoder based Collaborative Filtering}. One of the successful application is to consider the collaborative filtering from Autoencoder perspective.

\begin{figure}
\begin{center}
\begin{minipage}[t]{5.0cm}
\includegraphics[width=5.0cm]{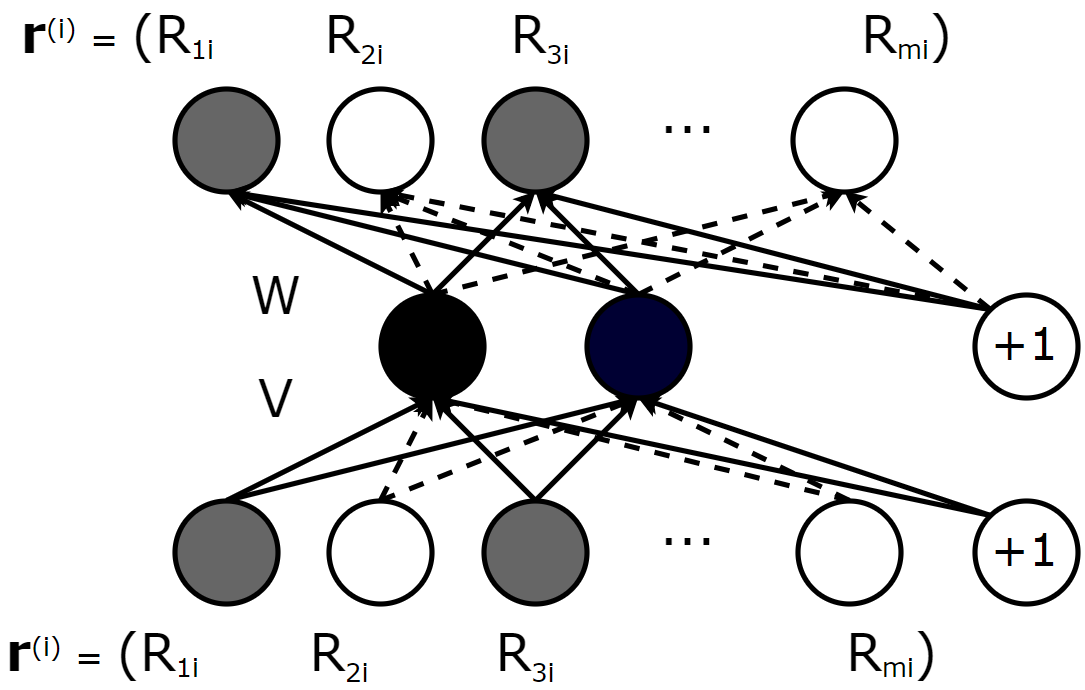}
\centering{(a)}
\end{minipage}
\begin{minipage}[t]{5.0cm}
\includegraphics[width=5.0cm]{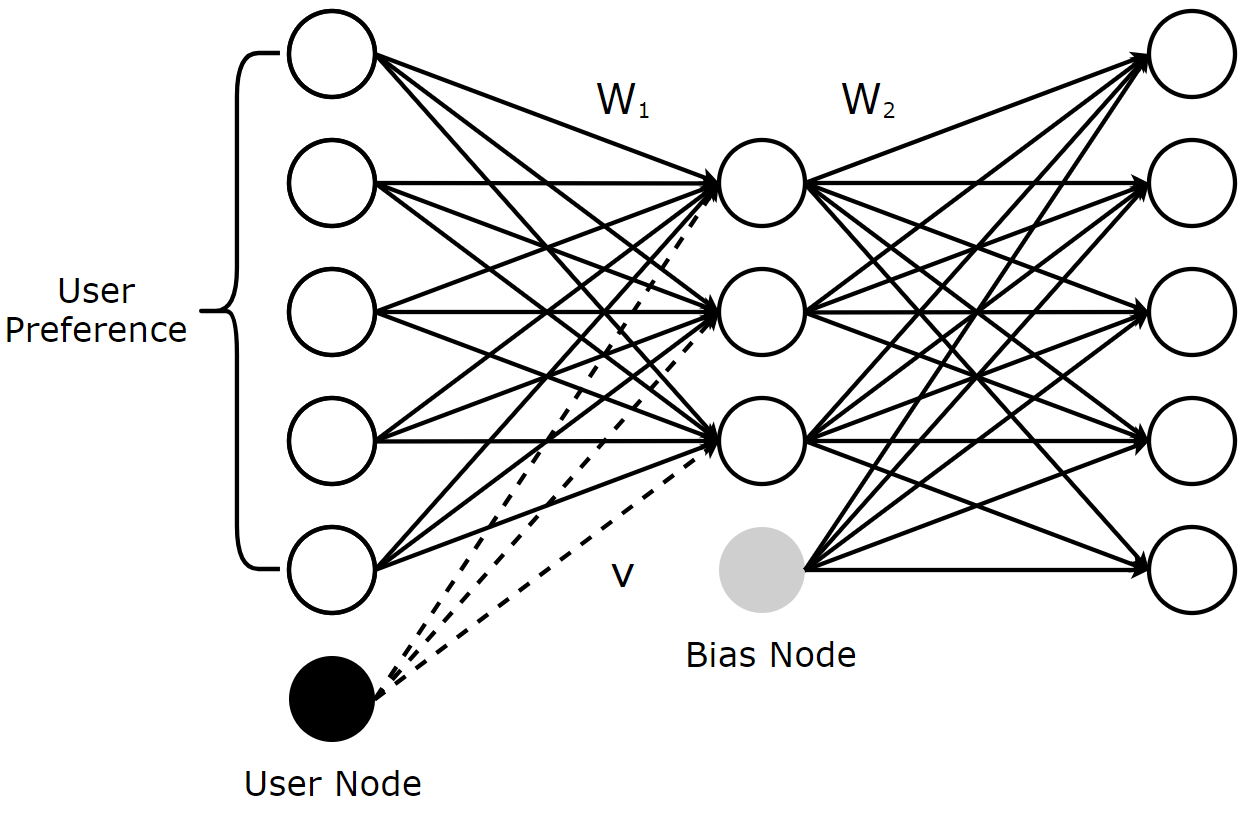}
\centering{(b)}
\end{minipage}
\begin{minipage}[t]{5.0cm}
\includegraphics[width=5.0cm]{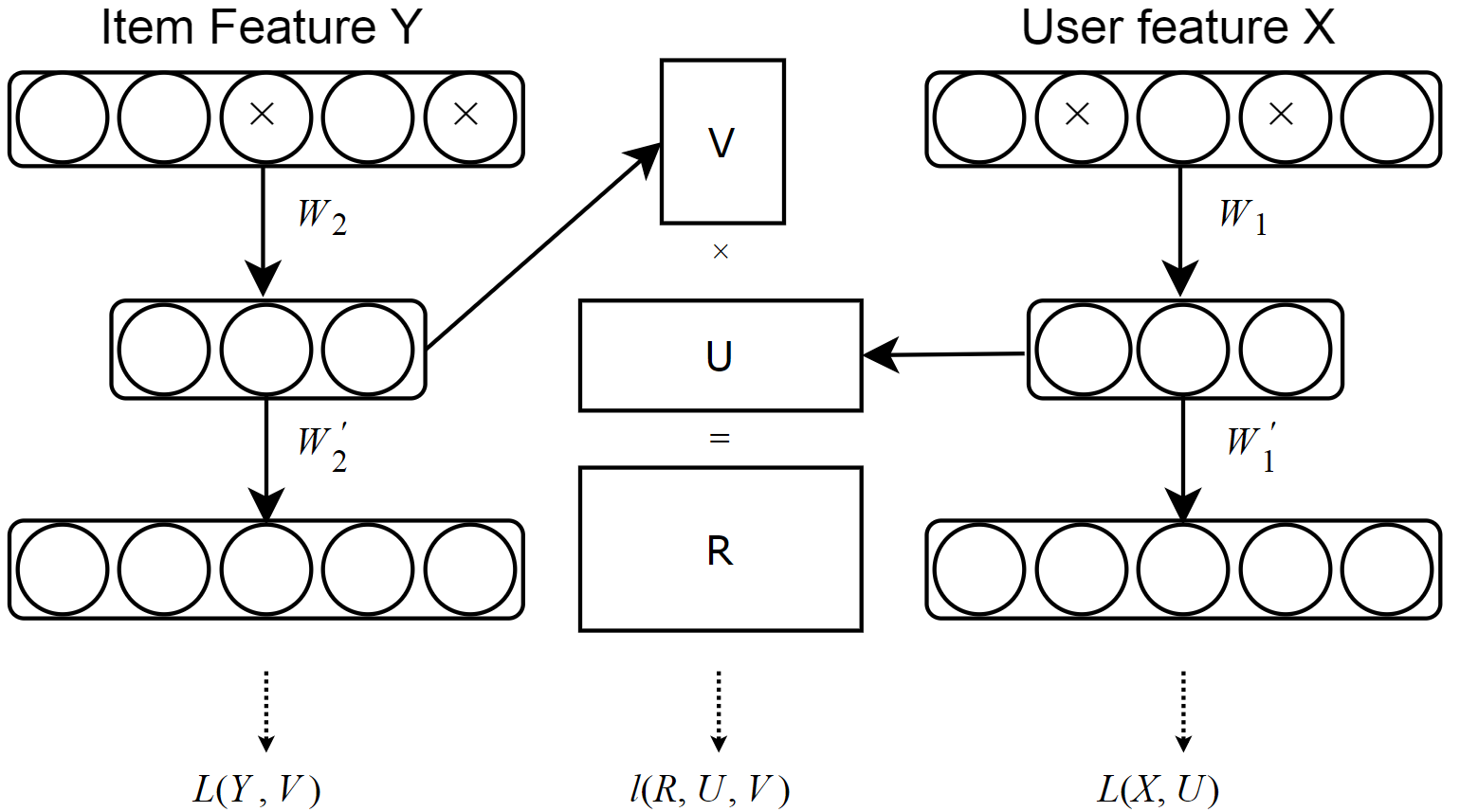}
\centering{(c)}
\end{minipage}
\caption{ Illustration of: (a) Item based AutoRec; (b) Collaborative denoising autoencoder; (c) Deep collaborative filtering framework.}
\label{fig:autoencodercf}
\end{center}
\vspace{-3mm}
\end{figure}

\textit{AutoRec}~\cite{sedhain2015autorec} takes user partial vectors $\textbf{r}^{(u)}$ or item partial vectors $\textbf{r}^{(i)}$ as input, and aims to reconstruct them in the output layer. Apparently, it has two variants: item-based AutoRec (I-AutoRec) and user-based AutoRec (U-AutoRec), corresponding to the two types of inputs. Here, we only introduce I-AutoRec, while U-AutoRec can be easily derived accordingly. Figure \ref{fig:autoencodercf}a illustrates the structure of I-AutoRec. Given input $\textbf{r}^{(i)}$, the reconstruction is:
$ h(\textbf{r}^{(i)};\theta) = f(W \cdot g(V \cdot \textbf{r}^{(i)} + \mu) + b)
$,
where $f(\cdot)$ and $g(\cdot)$ are the activation functions, parameter $\theta = \{W,V,\mu, b\}$. The objective function of I-AutoRec is formulated as follows:
\begin{equation}
    \underset{\theta}{argmin} \sum_{i=1}^N \parallel \textbf{r}^{(i)} - h(\textbf{r}^{(i)}; \theta)  \parallel_\mathcal{O}^2 + \lambda \cdot \textit{reg}
\end{equation}
Here $\parallel \cdot \parallel_\mathcal{O}^2$ means that it only considers observed ratings. The objective function can be optimized by resilient propagation (converges faster and produces comparable results) or L-BFGS (Limited-memory Broyden Fletcher Goldfarb Shanno algorithm). There are four important points about AutoRec that worth noticing before deployment: (1) I-AutoRec performs better than U-AutoRec, which may be due to the higher variance of user partially observed vectors. (2) Different combination of activation functions $f(\cdot)$ and $g(\cdot)$ will influence the performance considerably. (3) Increasing the hidden unit size moderately will improve the result as expanding the hidden layer dimensionality gives AutoRec more capacity to model the characteristics of the input. (4) Adding more layers to formulate a deep network can lead to slightly improvement.

\textit{CFN}~\cite{strub2016hybrid,strub2015collaborative} is an extension of AutoRec, and posses the following two advantages: (1) it deploys the denoising techniques, which makes CFN more robust; (2) it incorporates the side information such as user profiles and item descriptions to mitigate the sparsity and cold start influence. The input of CFN is also partial observed vectors, so it also has two variants: I-CFN and U-CFN, taking $\textbf{r}^{(i)}$ and $\textbf{r}^{(u)}$ as input respectively. Masking noise is imposed as a strong regularizer to better deal with missing elements (their values are zero). The authors introduced three widely used corruption approaches to corrupt the input: Gaussian noise, masking noise and salt-and-pepper noise. Further extension of CFN also incorporates side information. However, instead of just integrating side information in the first layer, CFN injects side information in every layer. Thus, the reconstruction becomes:
\begin{equation}
  h(\{\tilde{\textbf{r}}^{(i)}, \textbf{s}_i\}) = f(W_2  \cdot \{g(W_1 \cdot \{\textbf{r}^{(i)}, \textbf{s}_i\} + \mu), \textbf{s}_i\} + b)
\end{equation}
where $\textbf{s}_i$ is side information, $\{ \tilde{\textbf{r}}^{(i)}, \textbf{s}_i\}$ indicates the concatenation of $\tilde{\textbf{r}}^{(i)}$ and $\textbf{s}_i$. Incorporating side information improves the prediction accuracy, speeds up the training process and enables the model to be more robust.

\begin{table}[]
\centering
\caption{Summary of four autoencoder based recommendation models}
\label{table:autoencoder}
\begin{tabular}{|c|c|c|c|}
\hline
Vanilla/Denoising AE & Variational AE  & Contractive AE & Marginalized AE \\ \hline
  \begin{tabular}[c]{@{}c@{}} \cite{sedhain2015autorec,strub2016hybrid,strub2015collaborative,ouyang2014autoencoder,wu2016collaborative,wang2015collaborative}    \\
   \cite{ying2016collaborative,pana2017trust,jhamb2018attentive,wei2016collaborative, wei2017collaborative} \end{tabular} &       \cite{liang2018variational,li2016cvae,chen2018collective} &       \cite{zhang2017autosvd++}                  &     \cite{li2015deep}                                     \\ \hline
\end{tabular}
\end{table}

\textit{Collaborative Denoising Auto-Encoder (CDAE)}. The three models reviewed earlier are mainly designed for rating prediction, while CDAE~\cite{wu2016collaborative} is principally used for ranking prediction. The input of CDAE is user partially observed implicit feedback $\textbf{r}^{(u)}_{pref}$. The entry value is 1 if the user likes the movie, otherwise 0. It can also be regarded as a preference vector which reflects user's interests to items. Figure \ref{fig:autoencodercf}b illustrates the structure of CDAE. The input of CDAE is corrupted by Gaussian noise. The corrupted input $\tilde{\textbf{r}}^{(u)}_{pref}$ is drawn from a conditional Gaussian distribution $p(\tilde{\textbf{r}}^{(u)}_{pref} | \textbf{r}^{(u)}_{pref})$. The reconstruction is defined as:
\begin{equation}
  h(\tilde{\textbf{r}}^{(u)}_{pref}) = f(W_2  \cdot g(W_1 \cdot \tilde{\textbf{r}}^{(u)}_{pref} + V_u + b_1) + b_2)
\end{equation}
where $V_u \in \mathbb{R}^K $ denotes the weight matrix for user node (see figure \ref{fig:autoencodercf}b). This weight matrix is unique for each user and has significant influence on the model performance. Parameters of CDAE are also learned by minimizing the reconstruction error:
\begin{equation}
 \underset{W_1,W_2, V, b_1,b_2}{argmin}\frac{1}{M}\sum_{u=1}^M  \mathbf{E}_{p(\tilde{\textbf{r}}^{(u)}_{pref} | \textbf{r}^{(u)}_{pref})} [\ell (\tilde{\textbf{r}}^{(u)}_{pref},h(\tilde{\textbf{r}}^{(u)}_{pref}))] + \lambda \cdot \textit{reg}
\end{equation}
where the loss function $\ell(\cdot)$ can be square loss or logistic loss.

CDAE initially updates its parameters using SGD over all feedback. However, the authors argued that it is impractical to take all ratings into consideration in real world applications, so they proposed a negative sampling technique to sample a small subset from the negative set (items with which the user has not interacted), which reduces the time complexity substantially without degrading the ranking quality.

\textit{Muli-VAE and Multi-DAE}~\cite{liang2018variational} proposed a variant of varitional autoencoder for recommendation with implicit data, showing better performance than CDAE. The authors introduced a principled Bayesian inference approach for parameters estimation and show favorable results than commonly used likelihood functions.

To the extent of our knowledge, Autoencoder-based Collaborative Filtering (ACF)~\cite{ouyang2014autoencoder} is the first autoencoder based collaborative recommendation model. Instead of using the original partial observed vectors, it decomposes them by integer ratings. For example, if the rating score is integer in the range of [1-5], each $\textbf{r}^{(i)}$ will be divided into five partial vectors. Similar to AutoRec and CFN, the cost function of ACF aims at reducing the mean squared error.  However, there are two demerits of ACF: (1) it fails to deal with non-integer ratings; (2) the decomposition of partial observed vectors increases the sparseness of input data and leads to worse prediction accuracy.

\paratitle{Feature Representation Learning with Autoencoder}. Autoencoder is a class of powerful feature representation learning approach. As such, it can also be used in recommender systems to learn feature representations from user/item content features.

\begin{figure}
\begin{center}
\begin{minipage}[t]{14cm}
\includegraphics[width=14cm]{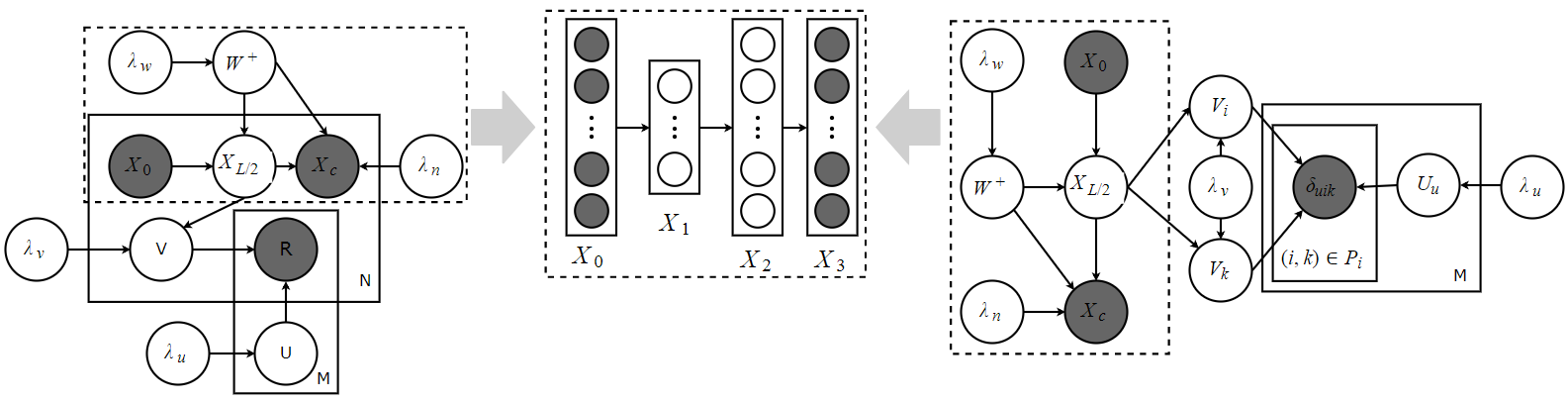}
\end{minipage}
\caption{ Graphical model of collaborative deep learning (left) and collaborative deep ranking (right).}
\label{fig:cdl}
\end{center}
\end{figure}

\textit{Collaborative Deep Learning (CDL)}.  CDL~\cite{wang2015collaborative} is a hierarchical Bayesian model which integrates stacked denoising autoencoder (SDAE) into probabilistic matrix factorization. To seamlessly combine deep learning and recommendation model, the authors proposed a general Bayesian deep learning framework~\cite{wang2016towards} consisting of two tightly hinged components: perception component (deep neural network) and task-specific component. Specifically, the perception component of CDL is a probabilistic interpretation of ordinal SDAE, and PMF acts as the task-specific component. This tight combination enables CDL to balance the influences of side information and interaction history. The generative process of CDL is as follows:

\begin{enumerate}
    \item For each layer $l$ of the SDAE: (a) For each column $n$ of weight matrix $W_l$, draw  $W_{l,*n} \sim \mathcal{N}(0,\lambda_w^{-1}\textbf{I}_{D_l})$; (b) Draw the bias vector $b_l \sim \mathcal{N}(0,\lambda_w^{-1}\textbf{I}_{D_l})$; (c) For each row $i$ of $X_l$, draw $X_{l,i*} \sim \mathcal{N}(\sigma(X_{l-1,i*}W_l + b_l),\lambda_s^{-1}\textbf{I}_{D_l})$.
    \item For each item $i$: (a) Draw a clean input $X_{c,i*} \sim \mathcal{N}(X_{L,i*},\lambda_n^{-1}\textbf{I}_{I_i})$; (b) Draw a latent offset vector $\epsilon_i \sim \mathcal{N}(0,\lambda_v^{-1}\textbf{I}_D)$ and set the latent item vector: $V_i = \epsilon_i + X_{\frac{L}{2},i*}^T$.
    \item Draw a latent user vector for each user $u$, $U_u \sim \mathcal{N}(0,\lambda_u^{-1}\textbf{I}_D)$.
    \item Draw a rating $r_{ui}$ for each user-item pair $(u, i)$, $r_{ui} \sim \mathcal{N}(U_u^T V_i,C_{ui}^{-1})$.
\end{enumerate}
where $W_l$ and $b_l$ are the weight matrix and biases vector for layer $l$, $X_l$ represents layer $l$. $\lambda_w$, $\lambda_s$, $\lambda_n$, $\lambda_v$, $\lambda_u$ are hyper-parameters, $C_{ui}$ is a confidence parameter for determining the confidence to observations~\cite{hu2008collaborative}. Figure \ref{fig:cdl}(left) illustrates the graphical model of CDL. The authors exploited an EM-style algorithm to learn the parameters. In each iteration, it updates $U$ and $V$ first, and then updates $W$ and $b$ by fixing $U$ and $V$. The authors also introduced a sampling-based algorithm~\cite{wang2016towards} to avoid the local optimum.

Before CDL, Wang et al.~\cite{wang2015relational} proposed a similar model, relational stacked denoising autoencoders (RSDAE), for tag recommendation. The difference of CDL and RSDAE is that RSDAE replaces the PMF with a relational information matrix. Another extension of CDL is collaborative variational autoencoder (CVAE)~\cite{li2016cvae}, which replaces the deep neural component of CDL with a variational autoencoder. CVAE learns probabilistic latent variables for content information and can easily incorporate multimedia (video, images) data sources.

\textit{Collaborative Deep Ranking (CDR)}. CDR~\cite{ying2016collaborative} is devised specifically in a pairwise framework for top-n recommendation. Some studies have demonstrated that pairwise model is more suitable for ranking lists generation~\cite{wu2016collaborative,ying2016collaborative,rendle2009bpr}. Experimental results also show that CDR outperforms CDL in terms of ranking prediction. Figure \ref{fig:cdl}(right) presents the structure of CDR. The first and second generative process steps of CDR are the same as CDL. The third and fourth steps are replaced by the following step:
\begin{itemize}
    \item For each user $u$: (a) Draw a latent user vector for $u$, $U_u \sim  \mathcal{N}(0,\lambda_u^{-1}\textbf{I}_D)$; (b) For each pair-wise preference $(i,j) \in P_i$, where $P_i = \{(i,j):r_{ui} - r_{uj} > 0\}$, draw the estimator, $\delta_{uij} \sim \mathcal{N}(U_u^TV_i - U_u^TV_j, C_{uij}^{-1}) $.
\end{itemize}
where $\delta_{uij} = r_{ui} - r_{uj}$ represents the pairwise relationship of user's preference on item $i$ and item $j$, $C_{uij}^{-1}$ is a confidence value which indicates how much user $u$ prefers item $i$ than item $j$. The optimization process is performed in the same manner as CDL.

\textit{Deep Collaborative Filtering Framework}. It is a general framework for unifying deep learning approaches with collaborative filtering model~\cite{li2015deep}. This framework makes it easily to utilize deep feature learning techniques to build hybrid collaborative models. The aforementioned work such as ~\cite{wang2015collaborative, wang2014improving, van2013deep} can be viewed as special cases of this general framework. Formally, the deep collaborative filtering framework is defined as follows:
\begin{equation}
\underset{U,V}{\arg \min}  \ell(R,U,V) + \beta(\parallel U \parallel_F^2 + \parallel V \parallel_F^2) + \gamma \mathcal{L}(X,U) + \delta \mathcal{L}(Y,V)
\end{equation}
where $\beta$, $\gamma$ and $\delta$ are trade-off parameters to balance the influences of these three components, $X$ and $Y$ are side information, $\ell(\cdot )$ is the loss of collaborative filtering model. $\mathcal{L}(X,U)$ and $\mathcal{L}(Y,V)$ act as hinges for connecting deep learning and collaborative models and link side information with latent factors. On top of this framework, the authors proposed the marginalized denoising autoencoder based collaborative filtering model (mDA-CF). Compared to CDL, mDA-CF explores a more computationally efficient variants of autoencoder: marginalized denoising autoencoder~\cite{chen2012marginalized}. It saves the computational costs for searching sufficient corrupted version of input by marginalizing out the corrupted input, which makes mDA-CF more scalable than CDL. In addition, mDA-CF embeds content information of items and users while CDL only considers the effects of item features.

\textit{AutoSVD++}~\cite{zhang2017autosvd++} makes use of contractive autoencoder~\cite{rifai2011contractive} to learn item feature representations, then integrates them into the classic recommendation model, SVD++~\cite{koren2008factorization}. The proposed model posses the following advantages: (1) compared to other autoencoders variants, contractive autoencoder captures the infinitesimal input variations; (2) it models the implicit feedback to further enhance the accuracy; (3) an efficient training algorithm is designed to reduce the training time.

\textit{HRCD}~\cite{wei2016collaborative, wei2017collaborative} is a hybrid collaborative model based on autoencoder and timeSVD++~\cite{koren2010collaborative}. It is a time-aware model which uses SDAE to learn item representations from raw features and aims at solving the cold item problem.

\subsection{Convolutional Neural Networks based Recommendation}
Convolution Neural Networks are powerful in processing unstructured multimedia data with convolution and pool operations. Most of the CNNs based recommendation models utilize CNNs for feature extraction.

\paratitle{Feature Representation Learning with CNNs}. CNNs can be used for feature representation learning from multiple sources such as image, text, audio, video, etc.

\textit{CNNs for Image Feature Extraction}. Wang et al.~\cite{Wang:2017:YIR:3038912.3052638} investigated the influences of visual features to Point-of-Interest (POI) recommendation, and proposed a visual content enhanced POI recommender system (VPOI). VPOI adopts CNNs to extract image features. The recommendation model is built on PMF by exploring the interactions between: (1) visual content and latent user factor; (2) visual content and latent location factor. Chu et al.~\cite{chu2017hybrid} exploited the effectiveness of visual information (e.g. images of food and furnishings of the restaurant) in restaurant recommendation. The visual features extracted by CNN joint with the text representation are input into MF, BPRMF and FM to test their performance. Results show that visual information improves the performance to some degree but not significant. He et al.~\cite{he2016vbpr} designed a visual Bayesian personalized ranking (VBPR) algorithm by incorporating visual features (learned via CNNs) into matrix factorization. He et al.~\cite{he2016ups} extended VBPR with exploring user's fashion awareness and the evolution of visual factors that user considers when selecting items. Yu et al.~\cite{yu2018aesthetic} proposed a coupled matrix and tensor factorization model for aesthetic-based clothing recommendation, in which CNNs is used to learn the images features and aesthetic features. Nguyen et al.~\cite{Nguyen2017} proposed a personalized tag recommendation model based on CNNs. It utilizes the convolutional and max-pooling layer to get visual features from patches of images. User information is injected for generating personalized recommendation. To optimize this network, the BPR objective is adopted to maximize the differences between the relevant and irrelevant tags. Lei et al.~\cite{lei2016comparative} proposed a comparative deep leaning model with CNNs for image recommendation. This network consists of two CNNs which are used for image representation learning and a MLP for user preferences modelling. It compares two images (one positive image user likes and one negative image user dislikes) against a user. The training data is made up of triplets: $t$ (user $U_t$, positive image $I^{+}_t$, negative image $I^{-}_t$). Assuming that the distance between user and positive image $D(\pi(U_t),\phi(I^{+}_t))$ should be closer than the distance between user and negative images $D(\pi(U_t),\phi(I^{-}_t))$, where $D(\cdot)$ is the distance metric (e.g. Euclidean distance). ConTagNet~\cite{rawat2016contagnet} is a context-aware tag recommender system. The image features are learned by CNNs. The context representations are processed by a two layers fully-connected feedforward neural network. The outputs of two neural networks are concatenated and fed into a softmax funcation to predict the probability of candidate tags.

\textit{CNNs for Text Feature Extraction}. DeepCoNN~\cite{Zheng:2017:JDM:3018661.3018665} adopts two parallel CNNs to model user behaviors and item properties from review texts. This model alleviates the sparsity problem and enhances the model interpretability by exploiting rich semantic representations of review texts with CNNs. It utilizes a word embedding technique to map the review texts into a lower-dimensional semantic space as well as keep the words sequences information. The extracted review representations then pass through a convolutional layer with different kernels, a max-pooling layer, and a full-connected layer consecutively. The output of the user network $x_u$ and item network $x_i$ are finally concatenated as the input of the prediction layer where the factorization machine is applied to capture their interactions for rating prediction. Catherine et al.~\cite{catherine2017transnets} mentioned that DeepCoNN only works well when the review text written by the target user for the target item is available at test time, which is unreasonable. As such, they extended it by introducing a latent layer to represent the target user-target-item pair. This model does not access the reviews during validation/test and can still remain good accuracy. Shen et al.~\cite{shen2016automatic} built an e-learning resources recommendation model. It uses CNNs to extract item features from text information of learning resources such as introduction and content of learning material, and follows the same procedure of \cite{van2013deep} to perform recommendation. ConvMF~\cite{kim2016convolutional} combines CNNs with PMF in a similar way as CDL. CDL uses autoencoder to learn the item feature representations, while ConvMF employs CNNs to learn high level item representations. The main advantage of ConvMF over CDL is that CNNs is able to capture more accurate contextual information of items via word embedding and convolutional kernels. Tuan et al.~\cite{tuan20173d} proposed using CNNs to learn feature representations form item content information (e.g., name, descriptions, identifier and category) to enhance the accuracy of session based recommendation.

\textit{CNNs for Audio and Video Feature Extraction}. Van et al.~\cite{van2013deep} proposed using CNNs to extract features from music signals. The convolutional kernels and pooling layers allow operations at multiple timescales. This content-based model can alleviate the cold start problem (music has not been consumed) of music recommendation. Lee et al.~\cite{lee2018collaborative} proposed extracting audio features with the prominent CNNs model ResNet. The recommendation is performed in the collaborative metric learning framework similar to CML.

\paratitle{CNNs based Collaborative filtering}. Directly applying CNNs to vanilla collaborative filtering is also viable. For example,  He et al.~\cite{he2018outer} proposed using CNNs to improve NCF and presented the ConvNCF. It uses outer product instead of dot product to model the user item interaction patterns. CNNs are applied over the result of outer product and could capture the high-order correlations among embeddings dimensions. Tang et al.~\cite{tang2018personalized} presented sequential recommendation (with user identifier) with CNNs, where two CNNs (hierarchical and vertical) are used to model the union-level sequential patterns and skip behaviors for sequence-aware recommendation.

\paratitle{Graph CNNs for Recommendation}. Graph convolutional Networks is a powerful tool for non-Eulcidean data such as: social networks, knowledge graphs, protein-interaction networks, etc~\cite{kipf2016semi}. Interactions in recommendation area can also be viewed as a such structured dataset (bipartite graph). Thus, it can also be applied to recommendation tasks. For example, Berg et al.~\cite{berg2017graph} proposed considering the recommendation problem as a link prediction task with graph CNNs. This framework makes it easy to integrate user/item side information such as social networks and item relationships into recommendation model. Ying et al.~\cite{ying2018graph} proposed using graph CNNs for recommendations in Pinterest\footnote{https://www.pinterest.com}. This model generates item embeddings from both graph structure as well item feature information with random walk and graph CNNs, and is suitable for very large-scale web recommender. The proposed model has been deployed in Pinterest to address a variety of real-world recommendation tasks.

\subsection{Recurrent Neural Networks based Recommendation}
RNNs are extremely suitable for sequential data processing. As such, it becomes a natural choice for dealing with the temporal dynamics of interactions and sequential patterns of user behaviours, as well as side information with sequential signals, such as texts, audio, etc.

\paratitle{Session-based Recommendation without User Identifier}.
In many real world applications or websites, the system usually does not bother users to log in so that it has no access to user's identifier and her long period consumption habits or long-term interests. However, the session or cookie mechanisms enables those systems to get user's short term preferences. This is a relatively unappreciated task in recommender systems due to the extreme sparsity of training data. Recent advancements have demonstrated the efficacy of RNNs in solving this issue~\cite{hidasi2015session, tan2016improved,wu2016personal}.

\begin{figure}
\begin{center}
\begin{minipage}[t]{4.8cm}
\includegraphics[width=4.8cm]{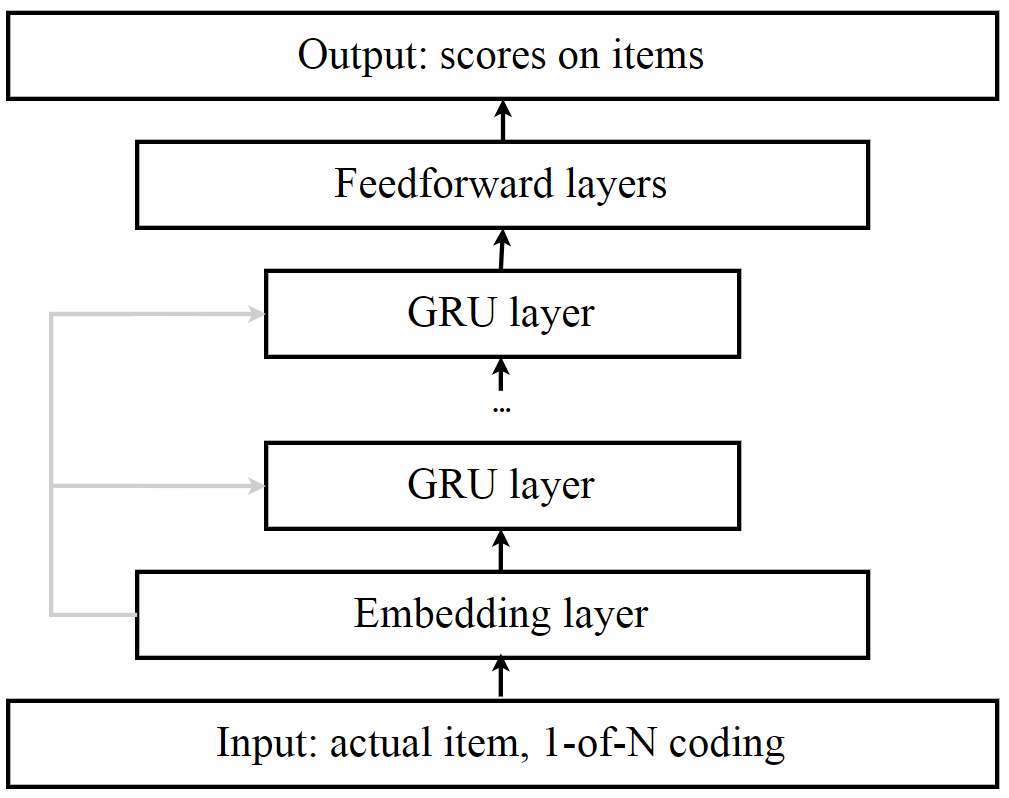}
\centering{(a)}
\end{minipage}
\begin{minipage}[t]{5.2cm}
\includegraphics[width=5.2cm]{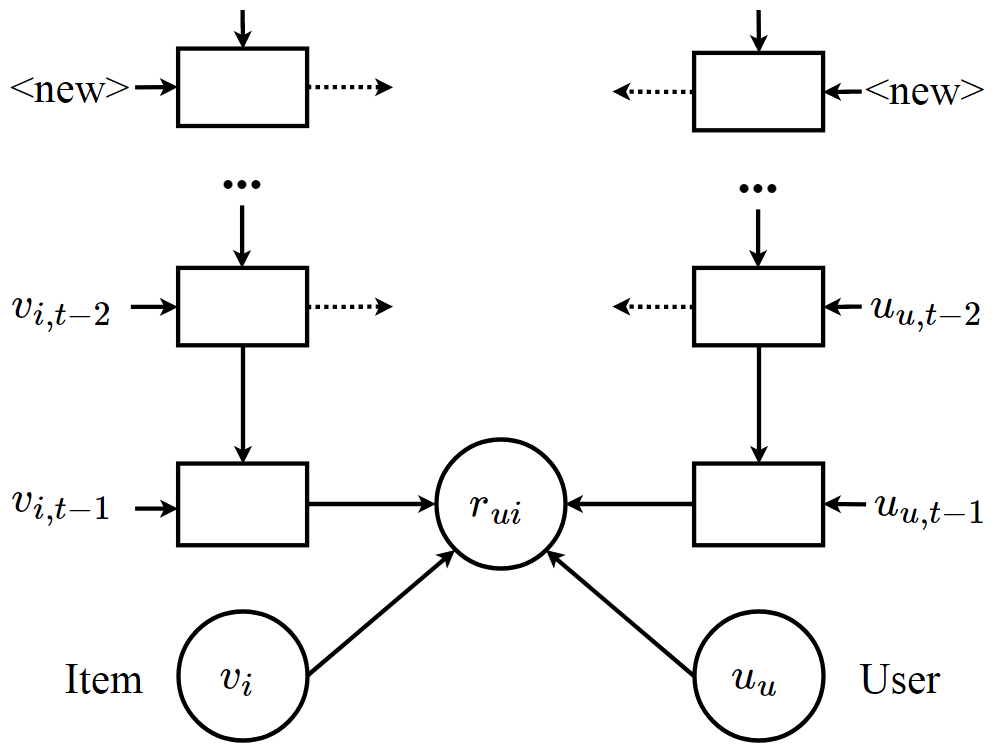}
\centering{(b)}
\end{minipage}
\begin{minipage}[t]{4.3cm}
\includegraphics[width=4.3cm]{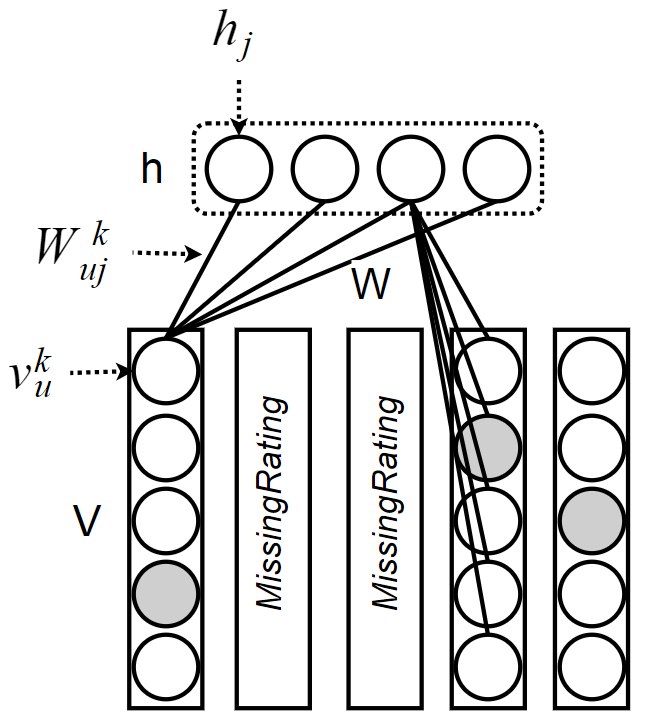}
\centering{(c)}
\end{minipage}
\caption{ Illustration of: (a) Session-based recommendation with RNN; (b) Recurrent recommender network; (c) Restricted Boltzmann Machine based Collaborative Filtering.}
\label{fig:gru4rec}
\end{center}
\end{figure}

\textit{GRU4Rec}. Hidasi et al.~\cite{hidasi2015session} proposed a session-based recommendation model, GRU4Rec, based GRU (shown in Figure \ref{fig:gru4rec}a). The input is the actual state of session with 1-of-$N$ encoding, where $N$ is the number of items. The coordinate will be 1 if the corresponding item is active in this session, otherwise 0. The output is the likelihood of being the next in the session for each item. To efficiently train the proposed framework, the authors proposed a session-parallel mini-batches algorithm and a sampling method for output. The ranking loss which is also coined TOP1 and has the following form:
\begin{equation}
\mathcal{L}_s = \frac{1}{S} \sum_{j=1}^S \sigma (\hat{r}_{sj} - \hat{r}_{si}) + \sigma(\hat{r}_{sj}^2)
\end{equation}
where $S$ is the sample size, $\hat{r}_{si}$ and $\hat{r}_{sj}$ are the scores on negative item $i$ and positive item $j$ at session $s$, $\sigma$ is the logistic sigmoid function. The last term is used as a regularization. Note that, BPR loss is also viable. A recent work~\cite{hidasi2017recurrent} found that the original TOP1 loss and BPR loss defined in~\cite{hidasi2015session} suffer from the gradient vanishing problem, as such, two novel loss functions: TOP1-max and BPR-max are proposed.

The follow-up work~\cite{tan2016improved} proposed several strategies to further improve this model: (1) augment the click sequences with sequence preprocessing and dropout regularization; (2) adapt to temporal changes by pre-training with full training data and fine-tuning the model with more recent click-sequences; (3) distillation the model with \textit{privileged information} with a teacher model; (4) using item embedding to decrease the number of parameters for faster computation.

Wu et al.~\cite{wu2016personal} designed a session-based recommendation model for real-world e-commerce website. It utilizes the basic RNNs to predict what user will buy next based on the click history. To minimize the computation costs, it only keeps a finite number of the latest states while collapsing the older states into a single history state. This method helps to balance the trade-off between computation costs and prediction accuracy. Quadrana et al.~\cite{quadrana2017personalizing} presented a hierarchical recurrent neural network for session-based recommendation. This model can deal with both session-aware recommendation when user identifiers are present.

The aforementioned three session-based models do not consider any side information. Two extensions~\cite{hidasi2016parallel,smirnova2017contextual} demonstrate that side information has effect on enhancing session recommendation quality. Hidasi et al.~\cite{hidasi2016parallel} introduced a parallel architecture for session-based recommendation which utilizes three GRUs to learn representations from identity one-hot vectors, image feature vectors and text feature vectors. The outputs of these three GRUs are weightedly concatenated and fed into a non-linear activation to predict the next items in that session. Smirnova et al.~\cite{smirnova2017contextual} proposed a context-aware session-based recommender system based on conditional RNNs. It injects context information into input and output layers. Experimental results of these two models suggest that models incorporated additional information outperform those solely based on historical interactions.

Despite the success of RNNs in session-based recommendation, Jannach et al.~\cite{Jannach:2017:RNN:3109859.3109872} indicated that simple neighbourhood approach could achieve same accuracy results as GRU4Rec. Combining the neighbourhood with RNNs methods can usually lead to best performance. This work suggests that some baselines in recent works are not well-justified and correctly evaluated. A more comprehensive discussion can be found in \cite{DBLP:journals/corr/abs-1803-09587}.

\paratitle{Sequential Recommendation with User Identifier}. Unlike session-based recommender where user identifiers are usually not present. The following studies deal with the sequential recommendation task with known user identifications.

\textit{Recurrent Recommender Network (RRN)}~\cite{wu2017recurrent} is a non-parametric recommendation model built on RNNs (shown in Figure \ref{fig:gru4rec}b). It is capable of modelling the seasonal evolution of items and changes of user preferences over time. RRN uses two LSTM networks as the building block to model dynamic user state $u_{ut}$ and item state $v_{it}$. In the meantime, considering the fixed properties such as user long-term interests and item static features, the model also incorporates the stationary latent attributes of user and item: $u_u$ and $v_i$. The predicted rating of item $j$ given by user $i$ at time $t$ is defined as:
\begin{equation}
\hat{r}_{ui|t} = f(u_{ut}, v_{it}, u_u, v_i)
\end{equation}
where $u_{ut}$ and $v_{it}$ are learned from LSTM, $u_u$ and $v_i$ are learned by the standard matrix factorization. The optimization is to minimize the square error between predicted and actual rating values.

Wu et al.~\cite{wu2016joint} further improved the RRNs model by modelling text reviews and ratings simultaneously. Unlike most text review enhanced recommendation models~\cite{Zheng:2017:JDM:3018661.3018665, seo2017representation}, this model aims to generate reviews with a character-level LSTM network with user and item latent states.  The review generation task can be viewed as an auxiliary task to facilitate rating prediction. This model is able to improve the rating prediction accuracy, but cannot generate coherent and readable review texts. NRT~\cite{li2017neural} which will be introduced in the following text can generate readable review tips. Jing et al.~\cite{jing2017neural} proposed a multi-task learning framework to simultaneously predict the returning time of users and recommend items. The returning time prediction is motivated by a survival analysis model designed for estimating the probability of survival of patients. The authors modified this model by using LSTM to estimate the returning time of costumers. The item recommendation is also performed via LSTM from user's past session actions. Unlike aforementioned session-based recommendations which focus on recommending in the same session, this model aims to provide inter-session recommendations. Li et al.~\cite{li2018learning} presented a behavior-intensive model for sequential recommendation. This model consists of two components: neural item embedding and discriminative behaviors learning. The latter part is made up of two LSTMs for session and preference behaviors learning respectively. Christakopoulou et al.~\cite{christakopoulou2018q} designed an interactive recommender with RNNs. The proposed framework aims to address two critical tasks in interactive recommender: ask and respond. RNNs are used to tackle both tasks: predict questions that the user might ask based on her recent behaviors(e.g, watch event) and predict the responses. Donkers et al.~\cite{donkers2017sequential} designed a novel type of Gated Recurrent Unit to explicit represent individual user for next item recommendation.

\paratitle{Feature Representation Learning with RNNs}. For side information with sequential patterns, using RNNs as the representation learning tool is an advisable choice.

Dai et al.~\cite{dai2016recurrent} presented a co-evolutionary latent model to capture the co-evolution nature of users' and items' latent features. The interactions between users and items play an important role in driving the changes of user preferences and item status. To model the historical interactions, the author proposed using RNNs to automatically learn representations of the influences from drift, evolution and co-evolution of user and item features.

Bansal et al.~\cite{bansal2016ask} proposed using GRUs to encode the text sequences into latent factor model. This hybrid model solves both warm-start and cold-start problems. Furthermore, the authors adopted a multi-task regularizer to prevent overfitting and alleviate the sparsity of training data. The main task is rating prediction while the auxiliary task is item meta-data (e.g. tags, genres) prediction.

Okura et al.~\cite{Shumpei2017} proposed using GRUs to learn more expressive aggregation for user browsing history (browsed news), and recommend news articles with latent factor model. The results show a significant improvement compared with the traditional word-based approach. The system has been fully deployed to online production services and serving over ten million unique users everyday.

Li et al.~\cite{li2017neural} presented a multitask learning framework, NRT, for predicting ratings as well as generating textual tips for users simultaneously. The generated tips provide concise suggestions and anticipate user's experience and feelings on certain products. The rating prediction task is modelled by non-linear layers over item and user latent factors $U \in \mathbb{R}^{k_u \times M}$, $V \in \mathbb{R}^{k_v \times M}$, where $k_u$ and $k_v$ (not necessarily equal) are latent factor dimensions for users and items. The predicted rating $r_{ui}$ and two latent factor matrices are fed into a GRU for tips generation. Here, $r_{ui}$ is used as context information to decide the sentiment of the generated tips. The multi-task learning framework enables the whole model to be trained efficiently in an end-to-end paradigm.

Song et al.~\cite{song2016multi} designed a temporal DSSM model which integrates RNNs into DSSM for recommendation. Based on traditional DSSM, TDSSM replace the left network with item static features, and the right network with two sub-networks to modelling user static features (with MLP) and user temporal features (with RNNs).

\subsection{Restricted Boltzmann Machine based Recommendation}

Salakhutdinov et al.~\cite{salakhutdinov2007restricted} proposed a restricted Boltzmann machine based recommender (shown in Figure \ref{fig:gru4rec}c). To the best of our knowledge, it is the first recommendation model that built on neural networks. The visible unit of RBM is limited to binary values, therefore, the rating score is represented in a one-hot vector to adapt to this restriction. For example, [0,0,0,1,0] represents that the user gives a rating score 4 to this item. Let $h_j, j=1,...,F$ denote the hidden units with fixed size $F$. Each user has a unique RBM with shared parameters. Suppose a user rated $m$ movies, the number of visible units is $m$, Let $X$ be a $K \times m$ matrix where $x_i^y=1$ if user $u$ rated movie $i$ as $y$ and $x_i^y=0$ otherwise. Then:
\begin{equation}
 p(v_i^y=1|h)= \frac{exp(b_i^y +\sum_{j=1}^Fh_jW_{ij}^y)}{\sum_{l=1}^Kexp(b_i^l+\sum_{j=1}^Fh_jW_{ij}^l)} \, \, \, \, \, , \, \, \, \, \,
 p(h_j=1|X)=\sigma(b_j+\sum_{i=1}^m\sum_{y=1}^Kx_i^yW_{ij}^y)
\end{equation}
where $W_{ij}^y$ represents the weight on the connection between the rating $y$ of movie $i$ and the hidden unit $j$, $b_i^y$ is the bias of rating $y$ for movie $i$, $b_j$ is the bias of hidden unit $j$. RBM is not tractable, but the parameters can be learned via the Contrastive Divergence (CD) algorithm~\cite{Goodfellow-et-al-2016}. The authors further proposed using a conditional RBM to incorporate the implicit feedback.  The essence here is that users implicitly tell their preferences by giving ratings, regardless of how they rate items.

The above RBM-CF  is user-based where a given user's rating is clamped on the visible layer. Similarity, we can easily design an item-based RBM-CF if we clamp a given item's rating on the visible layer. Georgiev et al.~\cite{georgiev2013non} proposed to combine the user-based and item-based RBM-CF in a unified framework. In the case, the visible units are determined both by user and item hidden units. Liu et al.~\cite{liu2015item} designed a hybrid RBM-CF which incorporates item features (item categories). This model is also based on conditional RBM. There are two differences between this hybrid model with the conditional RBM-CF with implicit feedback: (1) the conditional layer here is modelled with the binary item genres; (2) the conditional layer affects both the hidden layer and the visible layer with different connected weights.

\begin{table}[]
\centering
\caption{Categories of neural attention based recommendation models.}
\label{attrec}
\begin{tabular}{|c|c|}
\hline
\multicolumn{1}{|c|}{Vanilla Attention} & \multicolumn{1}{c|}{Co-Attention} \\ \hline
           \cite{chen2017acf,Tay:2018:LRM:3178876.3186154,jhamb2018attentive,gong2016hashtag,seo2017representation,wang2017dynamicattention,li2016hashtag,loyola2017modeling,liu2018stamp,ying2018sequential}                             &   \cite{zhou2017atrank,zhang2018dynamic,Tay:2018:MCN:3219819.3220086,zhanghashtag,hu2018leveraging}                                 \\ \hline
\end{tabular}
\end{table}

\subsection{Neural Attention based Recommendation}
Attention mechanism is motivated by human visual attention. For example, people only need to focus on specific parts of the visual inputs to understand or recognize them. Attention mechanism is capable of filtering out the uninformative features from raw inputs and reduce the side effects of noisy data. It is an intuitive but effective technique and has garnered considerable attention over the recent years across areas such as computer vision~\cite{ba2014multiple}, natural language processing~\cite{vaswani2017attention,luong2015effective} and speech recognition~\cite{chorowski2015attention,chorowski2014end}. Neural attention can not only used in conjunction with MLP, CNNs and RNNs, but also address some tasks independently~\cite{vaswani2017attention}. Integrating attention mechanism into RNNs enables the RNNs to process long and noisy inputs~\cite{chorowski2015attention}. Although LSTM can solve the long memory problem theoretically, it is still problematic when dealing with long-range dependencies. Attention mechanism provides a better solution and helps the network to better memorize inputs. Attention-based CNNs are capable of capturing the most informative elements of the inputs~\cite{seo2017representation}. By applying attention mechanism to recommender system, one could leverage attention mechanism to filter out uninformative content and select the most representative items~\cite{chen2017acf} while providing good interpretability. Although neural attention mechanism is not exactly a standalone deep neural technique, it is still worthwhile to discuss it separately due to its widespread use.

Attention model learns to attend to the input with attention scores. Calculating the attention scores lives at the heart of neural attention models. Based on the way for calculating the attention scores, we classify the neural attention models into (1) standard vanilla attention and (2) co-attention. Vanilla attention utilizes a parameterized context vector to learn to attend while co-attention is concerned with learning attention weights from two-sequences. Self-attention is a special case of co-attention.  Recent works~\cite{chen2017acf, gong2016hashtag,seo2017representation} demonstrate the capability of attention mechanism in enhancing recommendation performance. Table \ref{attrec} summarizes the attention based recommendation models.

\paratitle{Recommendation with Vanilla Attention}

Chen et al.~\cite{chen2017acf} proposed an attentive collaborative filtering model by introducing a two-level attention mechanism to latent factor model. It consists of item-level and component-level attention. The item-level attention is used to select the most representative items to characterize users. The component-level attention aims to capture the most informative features from multimedia auxiliary information for each user. Tay et al.~\cite{Tay:2018:LRM:3178876.3186154} proposed a memory-based attention for collaborative metric learning. It introduces a latent relation vector learned via attention to CML.  Jhamb et al.~\cite{jhamb2018attentive} proposed using attention mechanism to improve the performance of autoencoder based CF. Liu et al.~\cite{liu2018stamp} proposed a short-term attention and memory priority based model, in which both long and short term user interests are intergrated for session based recommendation. Ying et al.~\cite{ying2018sequential} proposed a hierarchical attention model for sequential recommendation. Two attention networks are used to model user long-term and short-term interests.

Introducing attention mechanism to RNNs could significantly improve their performance. Li et al.~\cite{li2016hashtag} proposed such an attention-based LSTM model for hashtag recommendation. This work takes the advantages of both RNNs and attention mechanism to capture the sequential property and recognize the informative words from microblog posts. Loyala et al.~\cite{loyola2017modeling} proposed an encoder-decoder architecture with attention for user session and intents modelling. This model consists of two RNNs and could capture the transition regularities in a more expressive way.

Vanilla attention can also work in conjunction with CNNs for recommender tasks. Gong et al.~\cite{gong2016hashtag} proposed an attention based CNNs system for hashtag recommendation in microblog. It treats hashtag recommendation as a multi-label classification problem. The proposed model consists of a global channel and a local attention channel. The global channel is made up of convolution filters and max-pooling layers. All words are encoded in the input of global channel. The local attention channel has an attention layer with given window size and threshold to select informative words (known as trigger words in this work). Hence, only trigger words are at play in the subsequent layers. In the follow-up work~\cite{seo2017representation}, Seo et al. made use of two neural networks same as \cite{gong2016hashtag} (without the last two layers) to learn feature representations from user and item review texts, and predict rating scores with dot product in the final layer. Wang et al.~\cite{wang2017dynamicattention} presented a combined model for article recommendation, in which CNNs is used to learn article representations and attention is utilized to deal with the diverse variance of editors's selection behavior.

\paratitle{Recommendation with Co-Attention} Zhang et al.~\cite{zhang2018dynamic} proposed a combined model, AttRec,  which improves  the sequential recommendation performance by capitalizing the strength of both self-attention and metric learning. It uses self-attention to learn user short-term intents from her recent interactions and takes the advantages of metric learning to learn more expressive user and item embemddings. Zhou et al.~\cite{zhou2017atrank} proposed using self-attention for user heterogeneous behaviour modelling.  Self-attention is simple yet effective mechanism and has shown superior performance than CNNs and RNNs in terms of sequential recommendation task. We believe that it has the capability to replace many complex neural models and more investigation is expected. Tay et al.~\cite{Tay:2018:MCN:3219819.3220086} proposed a review based recommendation system with multi-pointer co-attention. The co-attention enables the model to select information reviews via co-learning from both user and item reviews. Zhang et al.~\cite{zhanghashtag} proposed a co-atention based hashtag recommendation model that integrates both visual and textual information. Shi et al.~\cite{hu2018leveraging} proposed a neural co-attention model for personalized ranking task with meta-path.

\subsection{Neural AutoRegressive based Recommendation}
As mentioned above, RBM is not tractable, thus we usually use the Contrastive Divergence algorithm to approximate the log-likelihood gradient on the parameters~\cite{larochelle2011neural}, which also limits the usage of RBM-CF. The so-called Neural Autoregressive Distribution Estimator (NADE) is a tractable distribution estimator which provides a desirable alternative to RBM. Inspired by RBM-CF, Zheng et al.~\cite{Zheng:2016:NAA:3045390.3045472} proposed a NADE based collaborative filtering model (CF-NADE). CF-NADE models the distribution of user ratings.
Here, we present a detailed example to illustrate how the CF-NADE works. Suppose we have 4 movies: m1 (rating is 4), m2 (rating is 2), m3 (rating is 3) and m4 (rating is 5). The CF-NADE models the joint probability of the rating vector $r$ by the chain rule: $p(\textbf{r}) = \prod_{i=1}^D p(r_{m_{o_i}} | \textbf{r}_{m_{o_{<i}}})$,where $D$ is the number of items that the user has rated, $o$ is the $D$-tuple in the permutations of $(1,2,...,D)$, $m_i$ is the index of the $\textit{i}^{th}$ rated item, $r_{m_{o_i}}$ is the rating that the user gives to item $m_{o_i}$. More specifically, the procedure goes as follows: (1) the probability that the user gives  $m1$ 4-star conditioned on nothing; (2) the probability that the user gives $m2$ 2-star conditioned on giving $m1$ 4-star; (3) the probability that the user gives $m3$ 3-star conditioned on giving $m1$ 4-star and $m2$ 2-star; (4) the probability that the user gives $m4$ 5-star conditioned on giving $m1$ 4-star, $m2$ 2-star and $m3$ 3-star.



Ideally, the order of movies should follow the time-stamps of ratings. However, empirical study shows that random drawing also yields good performances. This model can be further extended to a deep model. In the follow-up paper, Zheng et al.~\cite{Zheng:2016:NAC:2988450.2988453} proposed incorporating implicit feedback to overcome the sparsity problem of rating matrix. Du et al.~\cite{du2016collaborative} further imporved this model with a user-item co-autoregressive approach, which ahieves better performance in both rating estimation and personalized ranking tasks.

\subsection{Deep Reinforcement Learning for Recommendation}
Most recommendation models consider the recommendation process as a static process, which makes it difficult to capture user's temporal intentions and to respond in a timely manner. In recent years, DRL has begun to garner attention~\cite{zhao2018deep,zhao2018recommendations,zheng2018drn,munemasadeep,choi2018reinforcement,wang2014exploration} in making personalized recommendation. Zhao et al.~\cite{zhao2018recommendations} proposed a DRL framework, DEERS, for recommendation with both negative and positive feedback in a sequential interaction setting.  Zhao et al.~\cite{zhao2018deep} explored the page-wise recommendation scenario with DRL, the proposed framework DeepPage is able to adaptively optimize a page of items based on user's real-time actions. Zheng et al.~\cite{zheng2018drn} proposed a news recommendation system, DRN, with DRL to tackle the following three challenges: (1) dynamic changes of news content and user preference; (2) incorporating return patterns (to the service) of users; (3) increase diversity of recommendations. Chen et al.~\cite{chen2018stabilizing} proposed a robust deep Q-learning algorithm to address the unstable reward estimation issue with two strategies: stratified
sampling replay and approximate regretted reward.  Choi et al.~\cite{choi2018reinforcement} proposed solving the cold-start problem with RL and bi-clustering. Munemasa et al~\cite{munemasadeep} proposed using DRL for stores recommendation.

Reinforcement Learning techniques such as contextual-bandit approach~\cite{li2010contextual} had shown superior recommendation performance in real-world applications. Deep neural networks increase the practicality of RL and make it possible to model various of extra information for designing real-time recommendation strategies.

\subsection{Adversarial Network based Recommendation}

IRGAN~\cite{wang2017irgan} is the first model which applies GAN to information retrieval area. Specifically, the authors demonstrated its capability in three information retrieval tasks, including: web search, item recommendation and question answering. In this survey, we mainly focus on how to use IRGAN to recommend items.

Firstly, we introduce the general framework of IRGAN. Traditional GAN consists of a discriminator and a generator. Likely, there are two schools of thinking in information retrieval, that is, generative retrieval and discriminative retrieval. Generative retrieval assumes that there is an underlying generative process between documents and queries, and retrieval tasks can be achieved by generating relevant document $d$ given a query $q$. Discriminative retrieval learns to predict the relevance score $r$ given labelled relevant query-document pairs. The aim of IRGAN is to combine these two thoughts into a unified model, and make them to play a minimax game like generator and discriminator in GAN. The generative retrieval aims to generate relevant documents similar to ground truth to fool the discriminative retrieval model.

Formally, let $p_{true}(d|q_n,r)$ refer to the user's relevance (preference) distribution. The generative retrieval model $p_{\theta}(d|q_n,r)$ tries to approximate the true relevance distribution. Discriminative retrieval $f_{\phi}(q,d)$ tries to distinguish between relevant documents and non-relevant documents. Similar to the objective function of GAN, the overall objective is formulated as follows:
\begin{equation}
J^{G^*, D^*} = \underset{\theta}{min} \ \underset{\phi}{max} \sum_{n=1}^N(\mathbb{E}_{d\sim p_{true}(d|q_n,r)}[log D(d|q_n)] + \mathbb{E}_{d\sim p_{\theta}(d|q_n,r)}[log(1- D(d|q_n))] )
\end{equation}
where $D(d|q_n) = \sigma(f_{\phi}(q,d))$, $\sigma$ represents the sigmoid function, $\theta$ and $\phi$ are the parameters for generative and discriminative retrieval respectively. Parameter $\theta$ and $\phi$ can be learned alternately with gradient descent.

The above objective equation is constructed for pointwise relevance estimation. In some specific tasks, it should be in pairwise paradigm to generate higher quality ranking lists. Here, suppose $p_{\theta}(d|q_n,r)$ is given by a softmax function:
\begin{equation}
p_{\theta}(d_i|q,r) = \frac{exp(g_{\theta}(q,d_i))}{\sum_{d_j} exp(g_{\theta}(q,d_j))}
\end{equation}
$g_{\theta}(q,d)$ is the chance of document $d$ being generated from query $q$. In real-word retrieval system, both $g_{\theta}(q,d)$ and $f_{\phi}(q,d)$ are task-specific. They can either have the same or different formulations. The authors modelled them with the same function for convenience, and define them as: $g_{\theta}(q,d) = s_{\theta}(q,d)$ and $f_{\phi}(q,d) = s_{\phi}(q,d)$. In the item recommendation scenario, the authors adopted the matrix factorization to formulate $s(\cdot)$. It can be substituted with other advanced models such as factorization machine or neural network.

He et al.~\cite{he2018adversarial} proposed an adversarial personalized ranking approach which enhances the Bayesian personalized ranking with adversarial training. It plays a minimax game between the original BPR objective and the adversary which add noises or permutations to maximize the BPR loss. Cai et al.~\cite{Cai2018GenerativeAN} proposed a GAN based representation learning approach for heterogeneous bibliographic network, which can effectively address the personalized citation recommendation task. Wang et al.~\cite{Wang:2018:NMS:3219819.3220004} proposed using GAN to generate negative samples for the memory network based streaming recommender. Experiments show that the proposed GAN based sampler could significantly improve the performance.

\subsection{Deep Hybrid Models for Recommendation}
With the good flexibility of deep neural networks, many neural building blocks can be intergrated to formalize more powerful and expressive models. Despite the abundant possible ways of combination, we suggest that the hybrid model should be reasonably and carefully designed for the specific tasks. Here, we summarize the existing models that has been proven to be effective in some application fields.


\paratitle{CNNs and Autoencoder}. Collaborative Knowledge Based Embedding (CKE)~\cite{zhang2016collaborative} combines CNNs with autoencoder for images feature extraction. CKE can be viewed as a further step of CDL. CDL only considers item text information (e.g. abstracts of articles and plots of movies), while CKE leverages structural content, textual content and visual content with different embedding techniques. Structural information includes the attributes of items and the relationships among items and users. CKE adopts the TransR~\cite{lin2015learning}, a heterogeneous network embedding method, for interpreting structural information. Similarly, CKE employs SDAE to learn feature representations from textual information. As for visual information, CKE adopts a stacked convolutional auto-encoders (SCAE). SCAE makes efficient use of convolution by replacing the fully-connected layers of SDAE with convolutional layers. The recommendation process is done in a probabilistic form similar to CDL.



\paratitle{CNNs and RNNs}. Lee et al.~\cite{lee2016quote} proposed a deep hybrid model with RNNs and CNNs for quotes recommendation. Quote recommendation is viewed as a task of generating a ranked list of quotes given the query texts or dialogues (each dialogue contains a sequence of tweets). It applies CNN sto learn significant local semantics from tweets and maps them to a distributional vectors. These distributional vectors are further processed by LSTM to compute the relevance of target quotes to the given tweet dialogues. The overall architecture is shown in Figure 12(a).

Zhang et al.~\cite{zhanghashtag} proposed a CNNs and RNNs based hybrid model for hashtag recommendation. Given a tweet with corresponding images, the authors utilized CNNs to extract features from images and LSTM to learn text features from tweets. Meanwhile, the authors proposed a co-attention mechanism to model the correlation influences and balance the contribution of texts and images.

Ebsesu et al.~\cite{ebesu2017neural} presented a neural citation network which integrates CNNs with RNNs in a encoder-decoder framework for citation recommendation. In this model, CNNs act as the encoder that captures the long-term dependencies from citation context. The RNNs work as a decoder which learns the probability of a word in the cited paper's title given all previous words together with representations attained by CNNs.

Chen et al.~\cite{Chen:2017:PKF:3077136.3080776} proposed an intergrated framework with CNNs and RNNs for personalized key frame (in videos) recommendation, in which CNNs are used to learn feature representations from key frame images and RNNs are used to process the textual features.

\paratitle{RNNs and Autoencoder}. The former mentioned collaborative deep learning model is lack of robustness and incapable of modelling the sequences of text information. Wang et al.~\cite{wang2016collaborative} further exploited integrating RNNs and denoising autoencoder to overcome this limitations. The authors first designed a generalization of RNNs named robust recurrent network. Based on the robust recurrent network, the authors proposed the hierarchical Bayesian recommendation model called CRAE. CRAE also consists of encoding and decoding parts, but it replaces feedforward neural layers with RNNs, which enables CRAE to capture the sequential information of item content information. Furthermore, the authors designed a wildcard denoising and a beta-pooling technique to prevent the model from overfitting.

\paratitle{RNNs with DRL}. Wang et al.~\cite{wang2018supervised} proposed combining supervised deep reinforcement learning wth RNNs for treatment recommendation.  The framework can learn the prescription policy from the indicator signal and evaluation signal. Experiments demonstrate that this system could infer and discover the optimal treatments automatically. We believe that this a valuable topic and benefits the social good.

\section{Future Research Directions and Open Issues}
Whilst existing works have established a solid foundation for deep recommender systems research, this section outlines several promising prospective research directions. We also elaborate on several open issues, which we believe is critical to the present state of the field.

\subsection{Joint Representation Learning from User and Item Content Information}
Making accurate recommendations requires deep understanding of item characteristics and user's actual demands and preferences~\cite{Leskovec:2015:NDR:2684822.2697044,adomavicius2005toward}. Naturally, this can be achieved by exploiting the abundant auxiliary information. For example, context information tailors services and products according to user's circumstances and surroundings~\cite{unger2016towards}, and mitigate cold start influence; Implicit feedback indicates users' implicit intention and is easier to collect while gathering explicit feedback is a resource-demanding task. Although existing works have investigated the efficacy of deep learning model in mining  user and item profiles~\cite{zhang2017autosvd++,lian2017cccfnet}, implicit feedback~\cite{ying2016collaborative,Zheng:2016:NAC:2988450.2988453,he2016vbpr,zhang2017autosvd++}, contextual information~\cite{unger2016towards,kim2016convolutional,rawat2016contagnet,twardowski2016modelling,ebesu2017neural}, and review texts~\cite{Zheng:2017:JDM:3018661.3018665,seo2017representation,wu2016joint,li2017neural} for recommendation, they do not utilize these various side information in a comprehensive manner and take the full advantages of the available data. Moreover, there are few works investigating users' footprints (e.g. Tweets or Facebook posts) from social media~\cite{hsieh2016immersive} and physical world (e.g. Internet of things)~\cite{yao2016things}. One can infer user's temporal interests or intentions from these side data resources while deep learning method is a desirable and powerful tool for integrating these additional information. The capability of deep learning in processing heterogeneous data sources also brings more opportunities in recommending diverse items with unstructured data such as textual, visual, audio and video features.


Additionally, feature engineering has not been fully studied in the recommendation research community, but it is essential and widely employed in industrial applications~\cite{covington2016deep,cheng2016wide}. However, most of the existing models require manually crafted and selected features, which is time-consuming and tedious. Deep neural network is a promising tool for automatic feature crafting by reducing manual intervention~\cite{shan2016deep}. There is also an added advantage of representation learning from free texts, images or data that exists in the `wild' without having to design intricate feature engineering pipelines. More intensive studies on deep feature engineering specific for recommender systems are expected to save human efforts as well as improve recommendation quality.

An interesting forward looking research problem is how to design neural architectures that best exploits the availability of other modes of data. One recent work potentially paving the way towards models of this nature is the Joint Representation Learning framework \cite{zhang2017joint}. Learning joint (possibly multi-modal representations) of user and items will likely become a next emerging trend in recommender systems research. To this end, a deep learning taking on this aspect would be how to design better inductive biases (hybrid neural architectures) in an end-to-end fashion. For example, reasoning over different modalities (text, images, interaction) data for better recommendation performance.


\subsection{Explainable Recommendation with Deep Learning}
A common interpretation is that deep neural networks are highly non-interpretable. As such, making explainable recommendations seem to be an uphill task. Along the same vein, it would be also natural to assume that big, complex neural models are just fitting the data with any \textit{true} understanding (see subsequent section on machine reasoning for recommendation). This is precisely why this direction is both exciting and also crucial. There are mainly two ways that explainable deep learning is important. The first, is to make explainable predictions to users, allowing them to understand the factors behind the network's recommendations (i.e., why was this item/service recommended?) \cite{xiao2017attentional,seo2017interpretable}. The second track is mainly focused on explain-ability to the practitioner, probing weights and activations to understand more about the model \cite{Tay:2018:LRM:3178876.3186154}.

As of today, attentional models \cite{Tay:2018:MCN:3219819.3220086,seo2017interpretable,xiao2017attentional} have more or less eased the non-interpretable concerns of neural models. If anything, attention models have instead led to greater extents of interpretability since the attention weights not only give insights about the inner workings of the model but are also able to provide explainable results to users. While this has been an existing direction of research `pre deep learning', attentional models are not only capable of enhancing performance but enjoys greater explainability. This further motivates the usage of deep learning for recommendation.

Notably, it is both intuitive and natural that a model's explainabiity and interpretability strongly relies on the application domain and usage of content information. For example \cite{seo2017interpretable,Tay:2018:MCN:3219819.3220086} mainly use reviews as a medium of interpretability (which reviews led to making which predictions). Many other mediums/modalities can be considered, such as image \cite{chen2018visually}.

To this end, a promising direction and next step would to be to design \textit{better} attentional mechanisms, possibly to the level of providing conversational or generative explanations (along the likes of \cite{li2017neural}). Given that models are already capable of highlighting what contributes to the decision, we believe that this is the next frontier.

\subsection{Going Deeper for Recommendation}
From former studies~\cite{he2017neural,he2017neural,wu2016collaborative,zhang2018neurec}, we found that the performance of most neural CF models plateaus at three to four layers. Going deeper has shown promising performance over shallow networks in many tasks~\cite{he2016deep,huang2017densely}, nonetheless, going deeper in the context of deep neural network based RS remains largely unclear. If going deeper give favorable results, how do we train the deep architecture? If not, what is the reason behind this? A possibility is to look into auxiliary losses at different layers in similar spirit to \cite{trinh2018learning} albeit hierarchically instead of sequentially. Another possibility is to vary layer-wise learning rates for each layer of the deep network or apply some residual strategies.

\subsection{Machine Reasoning for Recommendation}
There have been numerous recent advances in \textit{machine reasoning} in deep learning, often involving reasoning over natural language or visual input \cite{hudson2018compositional,santoro2017simple,xiong2016dynamic}. We believe that tasks like machine reading, reasoning, question answering or even visual reasoning will have big impacts on the field of recommender systems. These tasks are often glazed over, given that they seem completely arbitrary and irrelevant with respect to recommender systems. However, it is imperative that recommendater systems often requires reasoning over a single (or multiple) modalities (reviews, text, images, meta-data) which would eventually require borrowing (and adapting) techniques from these related fields. Fundamentally, recommendation and reasoning (e.g., question answering) are highly related in the sense that they are both information retrieval problems.

The single most impactful architectural innovation with neural architectures that are capable of machine reasoning is the key idea of attention \cite{vaswani2017attention,xiong2016dynamic}. Notably, this key intuition have already (and very recently) demonstrated effectiveness on several recommender problems. Tay et al. \cite{Tay:2018:MCN:3219819.3220086} proposed an co-attentive architecture for \textit{reasoning over reviews}, and showed that different recommendation domains have different `evidence aggregation' patterns. For interaction-only recommendation, similar reasoning architectures have utilized similar co-attentive mechanisms for reasoning over meta-paths \cite{hu2018leveraging}. To this end, a next frontier for recommender systems is possibly to adapt to situations that require multi-step inference and reasoning. A simple example would to reason over a user's social profile, purchases etc., reasoning over multiple modalities to recommend a product. All in all, we can expect that reasoning architectures to start to take the foreground in recommender system research.

\subsection{Cross Domain Recommendation with Deep Neural Networks}
Nowadays, many large companies offer diversified products or services to customers. For example, Google provides us with web searches, mobile applications and news services; We can buy books, electronics and clothes from Amazon. Single domain recommender system only focuses on one domain while ignores the user interests on other domains, which also exacerbates sparsity and cold start problems~\cite{Khan:2017:CDR:3101309.3073565}. Cross domain recommender system, which assists target domain recommendation with the knowledge learned from source domains, provides a desirable solution for these problems. One of the most widely studied topics in cross domain recommendation is transfer learning which aims to improve learning tasks in one domain by using knowledge transferred from other domains~\cite{fernandez2012cross,pan2010transfer}. Deep learning is well suited to transfer learning as it learn high-level abstractions that disentangle the variation of different domains. Several existing works~\cite{elkahky2015multi,lian2017cccfnet} indicate the efficacy of deep learning in catching the generalizations and differences across different domains and generating better recommendations on cross-domain platforms. Therefore, it is a promising but largely under-explored area where mores studies are expected.

\subsection{Deep Multi-Task Learning for Recommendation}
Multi-task learning has led to successes in many deep learning tasks, from computer vision to natural language processing~\cite{deng2014deep,collobert2008unified}. Among the reviewed studies, several works~\cite{jing2017neural, bansal2016ask, li2017neural, yi2016expanded} also applied multi-task learning to recommender system in a deep neural framework and achieved some improvements over single task learning. The advantages of applying deep neural network based multi-task learning are three-fold: (1) learning several tasks at a time can prevent overfitting by generalizing the shared hidden representations; (2) auxiliary task provides interpretable output for explaining the recommendation; (3) multi-task provides an implicit data augmentation for alleviating the sparsity problem. Multitask can be utilized in traditional recommender system~\cite{ning2010multi}, while deep learning enables them to be integrated in a tighter fashion. Apart from introducing side tasks, we can also deploy the multitask learning for cross domain recommendation with each specific task generating recommendation for each domain.




\subsection{Scalability of Deep Neural Networks for Recommendation}
The increasing data volumes in the big data era poses challenges to real-world applications. Consequently, scalability is critical to the usefulness of recommendation models in real-world systems, and the time complexity will also be a principal consideration for choosing models. Fortunately, deep learning has demonstrated to be very effective and promising in big data analytics~\cite{najafabadi2015deep} especially with the increase of GPU computation power. However, more future works should be studied on how to recommend efficiently by exploring the following problems: (1) incremental learning for non-stationary and streaming data such as large volume of incoming users and items; (2) computation efficiency for high-dimensional tensors and multimedia data sources; (3) balancing of the model complexity and scalability with the exponential growth of parameters. A promising area of research in this area involves knowledge distillation which have been explored in \cite{Tang:2018:RDL:3219819.3220021} for learning small/compact models for inference in recommender systems. The key idea is to train a smaller student model that absorbs knowledge from the large teacher model. Given that inference time is crucial for real time applications at a million/billion user scale, we believe that this is another promising direction which warrants further investigation. Another promising direction involves compression techniques~\cite{serra2017getting}. The high-dimensional input data can be compressed to compact embedding to reduce the space and computation time during model learning.

\subsection{The Field Needs Better, More Unified and Harder Evaluation}
Each time a new model is proposed, it is expected that the publication offers evaluation and comparisons against several baselines. The selection of baselines and datasets on most papers are seemingly arbitrary and authors generally have free reign over the choices of datasets/baselines. There are several issues with this.

Firstly, this creates an inconsistent reporting of scores, with each author reporting their own assortment of results. Till this day, there is seemingly on consensus on a general ranking of models (Notably, we acknowledge that the \textit{no free lunch theorem} exists). Occasionally, we find that results can be conflicting and relative positions change very frequently. For example, the scores of NCF in \cite{zheng2018mars} is relatively ranked very low as compared to the original paper that proposed the model \cite{he2017neural}. This makes the relative benchmark of new neural models extremely challenging. The question is how do we solve this? Looking into neighbouring fields (computer vision or natural language processing), this is indeed perplexing. Why is there no MNIST, ImageNet or SQuAD for recommender systems? As such, we believe that a suite of standardized evaluation datasets should be proposed.

We also note that datasets such as MovieLens are commonly used by many practioners in evaluating their models. However, test splits are often arbitrary (randomized). The second problem is that there is no control over the evaluation procedure. To this end, we urge the recommender systems community to follow the CV/NLP communities and establish a hidden/blinded test set in which prediction results can be only submitted via a web interface (such as Kaggle).

Finally, a third recurring problem is that there is no control over the difficulty of test samples in recommender system result. Is splitting by time the best? How do we know if test samples are either too trivial or impossible to infer? Without designing proper test sets, we argue that it is in fact hard to estimate and measure progress of the field. To this end, we believe that the field of recommender systems have a lot to learn from computer vision or NLP communities.



\section{Conclusion}

In this article, we provided an extensive review of the most notable works to date on deep learning based recommender systems. We proposed a classification scheme for organizing and clustering existing publications, and highlighted a bunch of influential research prototypes. We also discussed the  advantages/disadvantages of using deep learning techniques for recommendation tasks. Additionally, we detail some of the most pressing open problems and promising future extensions. Both deep learning and recommender systems are ongoing hot research topics in the recent decades. There are a large number of new developing techniques and emerging models each year. We hope this survey can provide readers with a comprehensive understanding towards the key aspects of this field, clarify the most notable advancements and shed some light on future studies.





\bibliographystyle{ACM-Reference-Format}
\bibliography{sigproc}